\documentclass[aip,graphicx,amsmath,amssymb,reprint]{revtex4-1}
\usepackage{graphicx}
\usepackage{dcolumn}
\usepackage{bm}

\usepackage[utf8]{inputenc}
\usepackage[T1]{fontenc}
\usepackage{mathptmx}
\usepackage{etoolbox}
\usepackage{mathtools}
\DeclareUnicodeCharacter{200E}{ }
\usepackage{multirow}
\begin{document}


\title{Probing the partition function for temperature-dependent potentials with nested sampling} 



\author{Lune Maillard}
\email[]{lune.maillard@insp.upmc.fr}
\affiliation{Sorbonne Université, CNRS, Institut des Nanosciences de Paris, INSP, F-75005 Paris, France}
\author{Philippe Depondt}
\affiliation{Sorbonne Université, CNRS, Institut des Nanosciences de Paris, INSP, F-75005 Paris, France}
\author{Fabio Finocchi}
\affiliation{Sorbonne Université, CNRS, Institut des Nanosciences de Paris, INSP, F-75005 Paris, France}
\author{Simon Huppert}
\affiliation{Sorbonne Université, CNRS, Institut des Nanosciences de Paris, INSP, F-75005 Paris, France}
\author{Thomas Plé}
\affiliation{Sorbonne Université, CNRS, Laboratoire de Chimie Théorique, LCT, F-75005 Paris, France}
\author{Julien Salomon}
\affiliation{Laboratoire Jacques-Louis Lions, Sorbonne Université and ANGE, INRIA Paris, France}
\author{Martino Trassinelli}
\affiliation{Sorbonne Université, CNRS, Institut des Nanosciences de Paris, INSP, F-75005 Paris, France}


\date{\today}

\begin{abstract}
Thermodynamic properties can be in principle derived from the partition function, which, in many-atom systems, is hard to evaluate as it involves a sum on the accessible microscopic states. Recently, the partition function has been computed via nested sampling, relying on Bayesian statistics, which is able to provide the density of states as a function of the energy in a single run, independently of the temperature. This appealing property is lost whenever the potential energy that appears in the partition function is temperature-dependent: for instance, mean-field effective potential energies or the quantum partition function in the path-integral formalism. For these cases, the nested sampling must be carried out at each temperature, which results in a massive increase of computational time. Here, we introduce and implement a new method, that is based on an extended partition function where the temperature is considered as an additional parameter to be sampled. The extended partition function can be evaluated by nested sampling in a single run, so to restore this highly desirable property even for temperature-dependent effective potential energies. We apply this original method to compute the quantum partition function for harmonic potentials and Lennard-Jones clusters at low temperatures and show that it outperforms the straightforward application of nested sampling for each temperature within several temperature ranges.

\end{abstract}


\maketitle 

\section{Introduction}\label{sec_intro}
The central quantity in statistical mechanics of many-atom systems is the partition function, from which all the thermodynamic properties can in principle be derived. For a system of $N$ classical particles, the partition function $Z$ is usually expressed as a multidimensional integral over $6N$ degrees of freedom, corresponding to the position vector $\boldsymbol{x}$ and the associated momenta $\boldsymbol{p}$ of the $N$ particles: 
\begin{equation}\label{eq_Z_intro}
Z(\beta)\propto\int d\boldsymbol{x}~d\boldsymbol{p}~\mathrm{e}^{-\beta H(\boldsymbol{x},\boldsymbol{p})} 
\end{equation}
with $\beta$ the inverse temperature and $H(\boldsymbol{x},\boldsymbol{p})$ the Hamiltonian. Furthermore, via the statistical weight $\exp(-\beta H)$, $Z$ is a function of the temperature. In practice, the computation of thermodynamic properties is very difficult as it implies sampling the high dimensional configurational space, which is computationally very demanding except for some very simple systems, subject for instance to a harmonic potential.

There is therefore a need for efficient numerical methods to tackle the problem of efficiently sampling the configurational space of a multi-atomic system. In the past, a variety of methods have been developed: parallel tempering \cite{hukushima_exchange_1996, calvo_phase_2000, predescu_thermodynamics_2005}, simulated annealing \cite{falcioni_biased_1999, lee_global_2000, hao_multiple_2015} and metadynamics \cite{micheletti_reconstructing_2004, quhe_path_2015} among others. In this work, we use the nested sampling algorithm \cite{skilling_nested_2004, ashton_nested_2022}. This algorithm, initially developed for data analysis applications, is a very general method that can be used to minimize the distance between models and experimental data: the model's parameter space is explored with an increasing likelihood constraint while simultaneously estimating the volume of parameter space in which the likelihood is above the constraint. This allows to better compare the relevance of different models, which standard least-square fit methods have difficulty doing. Bayesian model selection based on nested sampling has been applied in a variety of fields such as cosmology \cite{mukherjee_nested_2006}, biology \cite{mikelson_likelihood-free_2020}, microseismic events \cite{das_microseismic_2021} and many other fields, and its convergence has formally been proven \cite{evans_discussion_2007, chopin_properties_2010}.

In a different context, nested sampling can be used to explore the phase space of a physical system with a lowering energy constraint while simultaneously estimating the density of states (DOS), i.e., the number of states available at a given energy: the partition function and its derivatives can therefore be estimated in a very elegant manner. It has been applied to study the Potts model\cite{murray_nested_2005}, Lennard-Jones cluster \cite{partay_efficient_2010, griffiths_nested_2019}, transition path sampling \cite{bolhuis_nested_2018}, and the phase space of zirconium \cite{marchant_nested_2022} and water clusters \cite{szekeres_direct_2018}, among others. One of the advantage of this algorithm is that, for temperature-independent potentials, the thermodynamic quantities can be computed at all temperatures with only a single nested sampling exploration \cite{partay_efficient_2010}, contrary to the methods mentioned above.

However, this advantage is lost when the potential depends on the temperature, which is, for example, the case of mean-field theories \cite{kerman_mean-field_1981, goldenfeld_lectures_2018} where a physical variable is replaced by its temperature-dependent average\cite{goldenfeld_lectures_2018}, the Debye model \cite{kittel_introduction_2018} used for computing the specific heat of solids\cite{baggioli_explaining_2021}, or the path-integral formalism \cite{feynman_space-time_1948, markland_nuclear_2018} which takes into account the quantum nature of the nuclei. In this work, we consider and compare two methods to use nested sampling on such potentials: the \emph{direct method}, used in Ref. \onlinecite{szekeres_direct_2018}, which consists in performing several explorations in a pre-determined temperature range, one for each temperature that we wish to sample, and the \emph{extended partition function method} that allows to perform a single exploration in the case of a temperature-dependent potential and thus restores the "all-at-once" character of nested sampling. 

A particularly interesting field of application of these two methods is the path-integral formalism which is used to take into account nuclear quantum effects (NQEs) that can have a significant impact on the structure and dynamics of systems containing light element such as hydrogen\cite{markland_nuclear_2018}. The most direct manifestation of NQEs are isotope effects: indeed, for a classical system described by Eq. \eqref{eq_Z_intro}, the thermodynamic properties do not depend on the nuclear mass. Actually, isotope substitution can affect the properties of common compounds such as the melting temperature of water that increases by 4 K when replacing hydrogen with deuterium \cite{ceriotti_nuclear_2016, morrone_nuclear_2008}. In some cases, isotope substitution is known to even suppress some phase transitions entirely as in some ferroelectric materials \cite{muller_srti_1979, schaack_when_2023}. NQEs are also often relevant to phase transitions in high-pressure materials\cite{benoit_tunnelling_1998, bronstein_quantum_2016, schaack_quantum_2020}. To account for these NQEs in simulations in a numerically feasible manner, the path-integral formalism \cite{feynman_space-time_1948} consists in replacing the quantum nuclei by classical polymers made of $P$ replicas to mimic the quantum delocalization. This increases the numbers of degrees of freedom by a factor $P$, where $P$ is typically of the order of 10--100, making the computation of the thermodynamic quantities, such as the DOS, even more complex. In an effort to test new methods that can overcome this bottleneck, we test the two approaches based on nested sampling (the direct method and the extended partition function method) on two specific examples: the harmonic potential, for which analytical values are known, and atoms interacting via a Lennard-Jones potential. Another method that can be used to study isotope effects is thermodynamic integration \cite{ceriotti_efficient_2013, cheng_ab_2019} where the derivative of the free energy is integrated over the mass.

We first present the nested sampling algorithm in Section \ref{sec_ns} for the cases of temperature-independent and of temperature-dependent potentials for which we compare the direct and extended partition function methods. Next, in Section \ref{sec_PI_pres}, we recall the path-integral formalism. We then consider two applications: the quantum harmonic potential in Section \ref{sec_harmonic} and Lennard-Jones clusters in Section \ref{sec_LJ}. Finally, we conclude and give some perspectives for future work in Section \ref{sec_concl}.
 
\section{Nested sampling and the partition function}\label{sec_ns}
\subsection{Temperature-independent potential - General equations}
Let us consider a system composed of $N$ nuclei interacting via a potential $V(\boldsymbol{x})$, which depends only on the positions of the nuclei $\boldsymbol{x}$ and neither on their momenta nor on the temperature. For this system, the partition function $Z$ can be written as
\begin{equation}\label{eq_ev_vpos}
Z(\beta)=\frac{1}{N!~h^{3N}}\int  d\boldsymbol{x}~d\boldsymbol{p}~\exp\left(-\beta \left(\frac{1}{2}\sum_{i=1}^{3N}\frac{p_i^2}{m_i}+V(\boldsymbol{x})\right)\right)   
\end{equation}
where $\boldsymbol{p}=(p_1,...,p_{3N})$ is the vector of momenta, $(m_1,...,m_{3N})$ is the vector of masses, $\beta=\frac{1}{k_\textrm{B}T}$ the inverse temperature and $h$ the Planck constant. In this work, we consider the case where $m_i=m$ for all $1\leq i\leq 3N$ to simplify the notations but the generalization to different masses is straightforward. In this case, the partition function can be broken-down into two terms: a momentum-dependent one (kinetic contribution)
\begin{align}
Z_\textrm{k}(\beta)&=\frac{1}{N!~h^{3N}}\int d\boldsymbol{p}~\exp\left(-\beta \left(\frac{1}{2}\sum_{i=1}^{3N}\frac{p_i^2}{m}\right)\right)\nonumber\\
&=\left(\sqrt{\frac{2\pi m}{(N!)^2\beta h^2}}\right)^{3N}    
\end{align}
and a position-dependent one (configurational contribution)\cite{partay_nested_2021}
\begin{equation}\label{eq_pf_pos}
Z_c(\beta)= \int d\boldsymbol{x}~\mathrm{e}^{-\beta V(\boldsymbol{x})}=\int dE~\rho(E)~\mathrm{e}^{-\beta E},
\end{equation}
where $\rho(E)$ is the density of states, that is the number of microscopic configurations with energies between $E$ and $E+dE$:
\begin{equation}\label{eq_DOS_T_indep}
    \rho(E)=\int d\boldsymbol{x}~\delta(E-V(\boldsymbol{x})),
\end{equation}
and $\delta(\cdot)$ is the Dirac operator. The partition function $Z$ becomes the product of two contributions
\begin{equation}\label{eq_decomp_Z}
Z(\beta)=Z_\textrm{k}(\beta)\times Z_\textrm{c}(\beta).
\end{equation}

The relevant thermodynamic properties of the system can be deduced from $Z(\beta)$. As an example, the internal energy can be computed as
\begin{equation}\label{eq_U_decomp}
U=-\frac{\partial \log(Z)}{\partial \beta}=\frac{3N}{2}k_\textrm{B}T+\frac{\int dE~\rho(E) ~E~\mathrm{e}^{-\beta E}}{Z_\textrm{c}}=U_\textrm{k}+U_\textrm{c}, 
\end{equation}
where $U_\textrm{k}=3Nk_\textrm{B}T/2$ and $U_\textrm{c}=(\int dE~\rho(E) ~E~\mathrm{e}^{-\beta E})/Z_\textrm{c}$ are the momentum and position contributions, i.e., the average kinetic and potential energies, respectively. The heat capacity can be computed as
\begin{align}
C_\textrm{v}&=\frac{\partial U}{\partial T}\nonumber\\
&=\frac{3N}{2}k_\textrm{B}+\left(\frac{\int dE~\rho(E)~E^2~  \mathrm{e}^{-\beta E}}{k_\textrm{B} T^2 Z_\mathrm{c}}-\frac{\left(\int dE~\rho(E)~E~\mathrm{e}^{-\beta E}\right)^2}{k_\textrm{B} T^2Z_\mathrm{c}^2}\right)\nonumber \\
&=C_\textrm{v,k}+C_\textrm{v,c}, \label{eq_Cv_decomp}
\end{align}
where $C_\mathrm{v,k}=3Nk_\textrm{B}/2$ and 
\begin{equation} \label{eq_cvc_class}
    C_\mathrm{v,c}=\frac{\int dE~\rho(E)~E^2~  \mathrm{e}^{-\beta E}}{k_\textrm{B} T^2 Z_\mathrm{c}}-\frac{\left(\int dE~\rho(E)~E~\mathrm{e}^{-\beta E}\right)^2}{k_\textrm{B} T^2Z_\mathrm{c}^2}
\end{equation}
are the momentum and position contributions respectively.

In general, the position-dependent term $Z_\mathrm{c}$ does not have an analytical form. Thermodynamic properties such as $U$ and $C_\mathrm{v}$ thus have to be estimated numerically, generally using importance sampling of the phase space, for instance with some flavor of the Monte-Carlo Metropolis algorithm \cite{frenkel_understanding_2023}. Here, we use nested sampling, a method that has its origins in Bayes' theory of probabilities \cite{skilling_nested_2004}. 
\subsection{The nested sampling algorithm and estimation of the partition function}\label{sssec_ns_pres}
Let us recall the main steps of the nested sampling algorithm \cite{skilling_nested_2004, partay_efficient_2010} to sample the partition function:
\begin{enumerate}
    \item A relevant sampling space for positions is defined and $K$ points, called \emph{live points}, are uniformly sampled from this space. The integrals in Eqs. \eqref{eq_pf_pos} and \eqref{eq_decomp_Z} to \eqref{eq_cvc_class} will then be replaced by discrete sums over the sampling points.
    \item At each iteration $\ell$, the point $\boldsymbol{x}_{\mathrm{old}}$ associated with the highest energy is removed and replaced by a random point $\boldsymbol{x}_{\mathrm{new}}$ with an energy that is strictly lower: $V(\boldsymbol{x}_{\mathrm{new}})<V(\boldsymbol{x}_{\mathrm{old}})$. The partition function can then be estimated by the trapezoid rule\cite{ashton_nested_2022}: 
    \begin{equation}\label{eq_ev_calc_form_x_part}
       Z_\mathrm{c}(\beta)\approx\sum_{i=1}^\ell w_i \mathrm{e}^{-\beta E_i},
    \end{equation}
    where $E_\ell=V(\boldsymbol{x}_{\mathrm{old}})$ is the energy of the removed point at iteration $\ell$ and
    \begin{equation}\label{eq_weight}
     w_\ell=\frac{1}{2}\left(\mathrm{e}^{-\frac{\ell-1}{K}}-\mathrm{e}^{-\frac{\ell+1}{K}}\right)
    \end{equation}
    is an approximation of the DOS: $w_\ell\approx\rho(E)$. Equation \eqref{eq_ev_calc_form_x_part} is therefore a sum over the points removed at all previous iterations ($E_i$ and $w_i$ are respectively the energy of the removed points and the approximation of the DOS at iteration $i$, for $i\leq\ell$). More details are given in Appendix \ref{app_search}. 
    \item This procedure is repeated until the current contribution is small compared to previous contributions: $\log(w_\ell\mathrm{e}^{-\beta_\mathrm{s} E_\ell})-\log(c_{\mathrm{max}})<\delta$ with $c_{\mathrm{max}}=\max_{i\leq \ell}(w_i\mathrm{e}^{-\beta_\mathrm{s} E_i})$ at an inverse temperature $\beta_\mathrm{s}$ chosen by the user. This temperature $T_\mathrm{s}=1/(k_\textrm{B}\beta_\mathrm{s})$ needs to be small to ensure a good exploration of the phase space: it has to be inferior to the lowest temperature studied\cite{maillard_assessing_2023}. Here, we use $\delta=-10$, a value that satisfies our requirements. 
\end{enumerate}
At the end of the procedure, we thus have a collection of $(w_i,E_i)$ values, i.e., the DOS estimate $w_i$ at the energy $E_i$, from which we can compute an approximated partition function using Eq. \eqref{eq_ev_calc_form_x_part} with $\ell=n_{\mathrm{iter}}$, the number of iterations. We therefore sum over all the states that have been removed during the exploration, that is, until an energy minimum is reached. In the following, for simplicity, we will omit the bounds of the sum.

The advantage when using nested sampling on temperature-independent potentials is that $Z$ can be computed at all temperatures with only one exploration via nested sampling \cite{partay_efficient_2010} since the series of sampled $E_i$ does not depend on temperature but on the distribution of states (that is, the DOS) which is independent of $\beta$ as the potential $V(\boldsymbol{x})$ (see Eqs \eqref{eq_pf_pos} and \eqref{eq_DOS_T_indep}). This is also the case for the potential energy contribution to the internal energy and the heat capacity (see Eqs. \eqref{eq_U_decomp} and \eqref{eq_Cv_decomp}), which can be recovered as \cite{partay_efficient_2010}
\begin{equation}\label{eq_u_nf_class}
U_\mathrm{c}(\beta)=\frac{\sum_iw_iE_i\mathrm{e}^{-\beta E_i}}{Z_\mathrm{c}(\beta)}    
\end{equation}
and
\begin{equation}\label{eq_cv_nf_class}
C_\mathrm{v,c}(\beta)=\frac{\sum_iw_iE_i^2\mathrm{e}^{-\beta E_i}}{k_\textrm{B}T^2Z_\mathrm{c}(\beta)}-\frac{\left(\sum_iw_iE_i\mathrm{e}^{-\beta E_i}\right)^2}{k_\textrm{B}T^2Z_\mathrm{c}(\beta)^2},
\end{equation}
respectively.

All along the present work, we use the \texttt{nested\_fit} program \cite{trassinelli_nested_fit_2019, trassinelli_bayesian_2017, trassinelli_mean_2020, maillard_assessing_2023} which can be used both for model comparison in data analysis and for the estimation of partition functions in materials science. \texttt{Nested\_fit} has previously been used to study systems in which the potential energy does not depend on temperature, namely the harmonic potential \cite{maillard_assessing_2023} and Lennard-Jones clusters \cite{maillard_nested_2025}. In both cases, the program was able to recover the correct curves for the heat capacity. Here, we aim at studying the more complex case of temperature-dependent potentials, with a particular focus on quantum systems described within the path-integral framework.

\subsection{Temperature-dependent potential}
By temperature-dependent potential, we mean cases in which the partition function is written
\begin{equation}
    Z(\beta)=Z_\mathrm{k}(\beta)\times\int d\boldsymbol{x}~\mathrm{e}^{-\beta V(\boldsymbol{x},\beta)},
\end{equation}
where $Z_\mathrm{k}$ is the position-independent prefactor, that can be computed separately, and $V(\boldsymbol{x},\beta)$ is the effective potential depending on the positions $\boldsymbol{x}$ and inverse temperature $\beta$ (but not on the momenta). In analogy to Eq. \eqref{eq_pf_pos}, we define
\begin{equation}\label{eq_temp_dep_Z_c_gen}
    Z_\mathrm{c}(\beta)=\int d\boldsymbol{x}~\mathrm{e}^{-\beta V(\boldsymbol{x},\beta)}
\end{equation}
the position-dependent part of the partition function. In particular, from Eq. \eqref{eq_DOS_T_indep}, it follows that the DOS depends now on $T$:
\begin{equation}
    \rho(E,\beta)=\int d\boldsymbol{x}~\delta(E-V(\boldsymbol{x},\beta)).
\end{equation}

Similarly to Eqs. \eqref{eq_U_decomp} and \eqref{eq_Cv_decomp}, the internal energy and heat capacity can be recovered from the partition function. However, since the potential now depends on the temperature, the derivatives with respect to $\beta$ contain additional terms and the expressions are more involved:

\begin{align}\label{eq_u_temp_dep}
    U_\mathrm{c}&=\frac{\int d\boldsymbol{x} \left(V(\boldsymbol{x},\beta)+\beta\frac{\partial V(\boldsymbol{x},\beta)}{\partial \beta}\right)\mathrm{e}^{-\beta V(\boldsymbol{x},\beta)}}{Z_\mathrm{c}}\nonumber\\
    &=\left\langle V(\boldsymbol{x},\beta)+\beta\frac{\partial V(\boldsymbol{x},\beta)}{\partial \beta}\right\rangle,
\end{align}
\begin{align}\label{eq_cv_temp_dep}
    C_\mathrm{v,c}=&\frac{\int d\boldsymbol{x} \left(V(\boldsymbol{x},\beta)+\beta\frac{\partial V(\boldsymbol{x},\beta)}{\partial \beta}\right)^2\mathrm{e}^{-\beta V(\boldsymbol{x},\beta)}}{k_\textrm{B} T^2 Z_\mathrm{c}}\nonumber\\
    &-\frac{\left(\int d\boldsymbol{x} \left(V(\boldsymbol{x},\beta)+\beta\frac{\partial V(\boldsymbol{x},\beta)}{\partial \beta}\right)\mathrm{e}^{-\beta V(\boldsymbol{x},\beta)}\right)^2}{k_\textrm{B} T^2 Z_\mathrm{c}^2}\nonumber\\
    &-\frac{\int d\boldsymbol{x} \left(2\frac{\partial V(\boldsymbol{x},\beta)}{\partial \beta}+\beta\frac{\partial^2 V(\boldsymbol{x},\beta)}{\partial \beta^2}\right)\mathrm{e}^{-\beta V(\boldsymbol{x},\beta)}}{k_\textrm{B} T^2 Z_\mathrm{c}}\nonumber\\
    =&\frac{\left\langle \left(V(\boldsymbol{x},\beta)+\beta\frac{\partial V(\boldsymbol{x},\beta)}{\partial \beta}\right)^2\right\rangle-\left\langle V(\boldsymbol{x},\beta)+\beta\frac{\partial V(\boldsymbol{x},\beta)}{\partial \beta}\right\rangle^2}{k_\textrm{B} T^2}\nonumber\\
    &-\frac{\left\langle2\frac{\partial V(\boldsymbol{x},\beta)}{\partial \beta}+\beta\frac{\partial^2 V(\boldsymbol{x},\beta)}{\partial \beta^2}\right\rangle}{k_\textrm{B} T^2},
\end{align}
where $\langle\cdot\rangle$ denotes the average on the position coordinate $\boldsymbol{x}$ weighted by the probability distribution $\exp(-\beta V(\boldsymbol{x},\beta))/Z_\mathrm{c}$. We see that the first and second derivatives of the potential $V$ with respect to $\beta$ appear in the above expressions. Equations \eqref{eq_U_decomp} and \eqref{eq_Cv_decomp} for temperature-independent potentials can be recovered by taking $(\partial V(\boldsymbol{x},\beta))/(\partial \beta)=(\partial^2 V(\boldsymbol{x},\beta))/(\partial \beta^2)=0$.

Here, it is not straightforward to compute $Z$ at all temperatures in a single exploration as the $w_i$ and $E_i$ terms in Eq. \eqref{eq_ev_calc_form_x_part} depend on temperature. In the following sections, we specify the direct and the extended partition function for this case.
\subsubsection{Direct method}\label{sssec_dir_method_pres}
The direct method, introduced by Szekeres et al \cite{szekeres_direct_2018}, consists in performing one exploration per temperature of interest, using the nested sampling algorithm. For each value of the inverse temperature $\beta_j$, the temperature-independent potential $V(x,\beta_j)$ is sampled according to the procedure presented in Section \ref{sssec_ns_pres}. The partition function is thus recovered in the same manner as Eq. \eqref{eq_ev_calc_form_x_part}, for each inverse temperature $\beta_j$:
\begin{equation}
Z_\mathrm{c}(\beta_j)\approx\sum_iw_i\mathrm{e}^{-\beta_j E_i(\beta_j)},    
\end{equation}
where $E_i(\beta_j)$ are the energies sampled in the exploration performed at fixed inverse temperature $\beta_j$. The internal energy and heat capacity are computed by discretizing Eqs. \eqref{eq_u_temp_dep} and \eqref{eq_cv_temp_dep} using the fact that the average of a function $g$ depending on $\boldsymbol{x}$ and $\beta$ can be computed as
\begin{equation} \label{eq_nf_direct_average}
    \left\langle g(\boldsymbol{x},\beta_j)\right\rangle \approx \frac{\sum_i w_i~g(\boldsymbol{x}^{(i)},\beta_j)~\mathrm{e}^{-\beta_j E_i(\beta_j)}}{Z_\mathrm{c}(\beta_j)},
\end{equation}
where $\boldsymbol{x}^{(i)}$ is the position vector at iteration $i$. The full expressions of $U_\textrm{c}$ and $C_\textrm{v,c}$ are given in the supplementary material (Section I A).

\subsubsection{Extended partition function method}\label{sssec_ext_method_pres}
We now introduce a method allowing to obtain the same information for all temperatures in one exploration, as in the case of temperature-independent potential. The key idea of the method is to distinguish between the temperature $\beta$ at which the partition function is computed and the temperature which appears in the potential $V$, denoted hereafter by $\tilde{\beta}$, which we probe as an auxiliary parameter. Therefore, we sample both the positions $\boldsymbol{x}$ and the auxiliary temperature $\tilde{\beta}$, instead of just the positions as in the direct method, to eventually converge to the same physical results.

We want to compute the partition function from Eq. \eqref{eq_temp_dep_Z_c_gen}. To that end, we first define the \emph{extended partition function $\widetilde{Z}_\mathrm{c}(\beta)$} as
\begin{equation}
\widetilde{Z}_\mathrm{c}(\beta)=\int \int d\boldsymbol{x}~d\tilde{\beta}~\mathrm{e}^{-\beta V(\boldsymbol{x},\tilde{\beta})}.
\end{equation}
In this formula, the temperature parameter $\beta$ in $Z_\mathrm{c}(\beta)$ is split in two parameters which play distinct roles: the physical temperature $\beta$, at which the partition function is computed, and the auxiliary temperature $\tilde{\beta}$, which is a parameter used to sample the parameter space. The standard partition function $Z_\mathrm{c}(\beta)$ can be recovered from the extended partition function $\widetilde{Z}_\mathrm{c}(\beta)$ via
\begin{eqnarray}\label{eq_Z_1_run exact}
Z_\mathrm{c}(\beta)&=&\int \int d\boldsymbol{x}~d\tilde{\beta}~\delta(\beta-\tilde{\beta})~\mathrm{e}^{-\beta V(\boldsymbol{x},\tilde{\beta})}\\
&=&\left\langle\delta(\beta-\tilde{\beta})\right\rangle_{\tilde{\beta}}\widetilde{Z}_\mathrm{c}(\beta),    
\end{eqnarray}
where $\left\langle\delta(\beta-\tilde{\beta})\right\rangle_{\tilde{\beta}}$ is the average, according to the probability distribution $\exp(-\beta V(\boldsymbol{x},\tilde{\beta}))/\widetilde{Z}_\mathrm{c}(\beta)$, of the Dirac delta-function $\delta(\beta-\tilde{\beta})$ in the extended $(\boldsymbol{x},\tilde{\beta})$ space. In the above expression, $Z_\textrm{c}(\beta)$ corresponds to the partition function at inverse temperature $\beta$ of the potential $V(\boldsymbol{x},\tilde{\beta})$, from which the delta-function only selects the cases where $\beta=\tilde{\beta}$. As in $\widetilde{Z}_\mathrm{c}(\beta)$, the potential $V$ does not depend on the inverse temperature $\beta$ but on the auxiliary inverse temperature $\tilde{\beta}$: it is therefore possible to compute $Z_\mathrm{c}(\beta)$ at all temperatures in a single exploration, as in the classical case, but at the cost of adding an extra parameter in the exploration, namely $\tilde{\beta}$. This method is thus referred to as the \emph{extended partition function} method or \emph{extended} method in short. 

Ideally, to compute $Z_\mathrm{c}(\beta)$ at temperature $\beta$, we would like to use only the values of $\tilde{\beta}$ for which $\tilde{\beta}=\beta$. However, in practice, because of the use of discrete sampling, a true delta-function is too restrictive: the probability to sample points for which $\tilde{\beta}=\beta$ is null. We therefore use a function $f(t;\alpha)$, depending on a parameter $\alpha$, that approximates the delta function \cite{kanwal_generalized_2004} in the $\alpha\to\infty$ limit, i.e., $\lim_{\alpha\to\infty}f(t;\alpha)=\delta(t)$. As a result, instead of selecting the values where $\tilde{\beta}=\beta$, we choose the points sampled at a temperature $\tilde{\beta}$ where $\tilde{\beta}\approx\beta$: a weight is given to each value of $\tilde{\beta}$ through the function $f$ and used to compute $Z_\mathrm{c}(\beta)$:
\begin{equation}\label{eq_Z_1_run approx}
Z_\mathrm{c}(\beta)=\left\langle\delta(\beta-\tilde{\beta})\right\rangle_{\tilde{\beta}}\widetilde{Z}_\mathrm{c}(\beta)\approx \left\langle f(\beta-\tilde{\beta};\alpha)\right\rangle_{\tilde{\beta}}\widetilde{Z}_\textrm{c}(\beta).
\end{equation}
In the following, the double-brackets $\left\langle\left\langle \cdot\right\rangle\right\rangle_\alpha$ will denote
\begin{equation}\label{eq_av_g_extended_alpha}
    \left\langle\left\langle g(\boldsymbol{x},\beta)\right\rangle\right\rangle_\alpha = \left\langle f(\beta-\tilde{\beta};\alpha) g(\boldsymbol{x},\tilde{\beta})\right\rangle_{\tilde{\beta}}  
\end{equation}
the average of a function $g$ according to the probability distribution $\exp(-\beta V(\boldsymbol{x},\tilde{\beta}))/\widetilde{Z}_\mathrm{c}(\beta)$ using $f$ as the delta-function approximation with parameter $\alpha$. The average also depends on the choice of the prior on $\tilde{\beta}$, which will be discussed in more details in Section \ref{sssec_extended_method_sampling} for the path-integral formalism. Equation \eqref{eq_av_g_extended_alpha} can be discretized using 
\begin{equation}\label{eq_nf_extended_average}
    \left\langle\left\langle g(\boldsymbol{x},\beta)\right\rangle\right\rangle_\alpha \approx \frac{\sum_i w_i g(\boldsymbol{x}^{(i)}, \tilde{\beta}_i) f(\beta-\tilde{\beta}_i;\alpha)\mathrm{e}^{-\beta E_i}}{\widetilde{Z}_\textrm{c}(\beta)}
\end{equation}
with $w_i$ as in Eq. \eqref{eq_weight} and $\boldsymbol{x}^{(i)}$ the position vector, $\tilde{\beta}_i$ and $E_i$ the inverse temperature and energy sampled at iteration $i$. The term $E_i=E(\boldsymbol{x}^{(i)},\tilde{\beta}_i)$ is computed at inverse temperature $\tilde{\beta}_i$ for the position vector $\boldsymbol{x}^{(i)}$. Note that, in practice, increasing the value of $\alpha$ , while providing a better approximation of the delta function, results in a smaller window and therefore to small amounts of sampling points being used to compute the thermodynamic properties at each temperature. On the contrary, small values of $\alpha$ use more sampling points but those points are sampled on a wider temperature range. The impact of the choice of $\alpha$ and therefore of $f(\beta-\tilde{\beta};\alpha)$ in Eq. \eqref{eq_av_g_extended_alpha}  will be discussed in Sections \ref{sec_harmonic} (harmonic potentials) and \ref{sec_LJ} (Lennard-Jones clusters).

Similarly, the internal energy and heat capacity are computed as (see Eqs. \eqref{eq_u_temp_dep} and \eqref{eq_cv_temp_dep})
\begin{equation} \label{eq_u_ext_method_exact}
U_\textrm{c}(\beta)\approx\frac{\widetilde{Z}_{\beta}}{Z_\textrm{c}(\beta)}\left\langle\left\langle V(\boldsymbol{x},\beta)+\beta\frac{\partial V(\boldsymbol{x},\beta)}{\partial \beta}\right\rangle\right\rangle_{\alpha}    
\end{equation}
and
\begin{align} \label{eq_cv_ext_method_exact}
C_\textrm{v,c}(\beta)\approx&\frac{\widetilde{Z}_\textrm{c}(\beta)}{k_\textrm{B} T^2 Z_\textrm{c}(\beta)}\left\langle\left\langle \left(V(\boldsymbol{x},\beta)+\beta\frac{\partial V(\boldsymbol{x},\beta)}{\partial \beta}\right)^2\right\rangle\right\rangle_{\alpha}\nonumber\\
-&\frac{\widetilde{Z}_\textrm{c}(\beta)^2}{k_\textrm{B} T^2 Z_\textrm{c}(\beta)^2}\left(\left\langle\left\langle \left(V(\boldsymbol{x},\beta)+\beta\frac{\partial V(\boldsymbol{x},\beta)}{\partial \beta}\right)\right\rangle\right\rangle_{\alpha}\right)^2\nonumber \\
-& \frac{\widetilde{Z}(\beta)}{k_\textrm{B}T^2 Z_\textrm{c}(\beta)}\left\langle\left\langle \left(2\frac{\partial V(\boldsymbol{x},\beta)}{\partial \beta}+\beta\frac{\partial^2 V(\boldsymbol{x},\beta)}{\partial \beta^2}\right)\right\rangle\right\rangle_{\alpha},    
\end{align}
respectively. The full expressions of $U_\textrm{c}$ and $C_\textrm{v,c}$ are given in the supplementary material (Section I B).

From a computational point of view, with the extended method, the series of $(w_i,E_i,\tilde{\beta}_i)$ is sampled once to compute the thermodynamic properties in a pre-determined temperature range, in contrast with the direct method where the sampling must be carried out separately for each temperature. While the sampling is more expensive for the extended method, we anticipate that the total number of required samples is smaller than for the direct method. 

In this work, we consider two choices for the function $f$ (both normalized to unity) among the many possible approximations of the Dirac delta-function:
\begin{itemize}
    \item the rectangular function 
    \begin{equation}
    f_\textrm{r}(\beta-\tilde{\beta};\alpha)=\begin{cases}
        \alpha/2 & \mbox{ if } |\beta-\tilde{\beta}|<1/\alpha,\\
        0 & \mbox{ otherwise};
    \end{cases}   
    \end{equation}
    $f_\textrm{r}$ selects all sampled points for which $\tilde{\beta}$ is $1/\alpha$-close to $\beta$ giving them all the same weight.
    \item the Gaussian function 
    \begin{equation}
    f_\textrm{G}(\beta-\tilde{\beta};\alpha)=\frac{\alpha}{\sqrt{\pi}}\mathrm{e}^{-\alpha^2 (\beta-\tilde{\beta})^2};   
    \end{equation}
    $f_\textrm{G}$ gives a normal weight to all sampled points: points for which $\tilde{\beta}$ is distant from $\beta$ of more than $3\frac{1}{\sqrt{2}\alpha}$ (three standard deviations) will have a very small weight and participate little to $Z_\textrm{c}^P(\beta)$.
\end{itemize}
In both case, $\alpha$ has the dimension of an energy.
\section{The quantum partition function in the path-integral formalism}\label{sec_PI_pres}
\subsection{General equations}
In this section, we present the specific form of the partition function for the particular case of a temperature-dependent potential resulting from the treatment of NQEs via Feynman's path-integral formalism \cite{feynman_quantum_2010}. The quantum nuclei are represented by classical closed polymers composed of $P$ replicas. When $P\to\infty$, the probability distribution for the replicas converges toward the exact quantum distribution. The partition function $Z^P$ of this new system for $N$ atoms in three dimensions and $P$ replicas is \cite{tuckerman_statistical_2010, szekeres_direct_2018}
\begin{equation}\label{eq_quantum_partition_function}
Z^P(\beta)=\left(\frac{P m}{2\pi\beta\hbar^2}\right)^{3NP/2}\idotsint d\boldsymbol{x}_1~...~d\boldsymbol{x}_P\mathrm{e}^{-\beta V_P}   
\end{equation}
with the potential
\begin{equation}
V_P(\boldsymbol{x}_1,...,\boldsymbol{x}_P,\beta)=\underbrace{\frac{mP}{2\hbar^2\beta^2}\sum_{i=1}^P(\boldsymbol{x}_i-\boldsymbol{x}_{i+1})^2}_{Q(\boldsymbol{x},\beta)}+\underbrace{\frac{1}{P}\sum_{i=1}^P V(\boldsymbol{x}_i)}_{\bar{V}(\boldsymbol{x})}~~~\label{eq_potentiel_PI}
\end{equation}
The term $\bar{V}$ is the interaction potential $V$ of the quantum nuclei averaged over all replicas and $Q$ is a term that derives from the quantum kinetic operator and corresponds to an harmonic interaction between the replicas.  We denote
\begin{equation}\label{eq_vect_x_vp}
\boldsymbol{x}_i=(x_{i,1},y_{i,1},z_{i,1},...,x_{i,N},y_{i,N},z_{i,N})    
\end{equation}
the position vector of the $i$-th replica of the system and $\boldsymbol{x}=(\boldsymbol{x}_1,...,\boldsymbol{x}_P)$ the full position vector. In that case, we have
\begin{equation}\label{eq_V_der_1}
    \frac{\partial V_P(\boldsymbol{x},\beta)}{\partial \beta}=-\frac{2}{\beta}Q(\boldsymbol{x},\beta)
\end{equation}
and
\begin{equation}\label{eq_V_der_2}
    \frac{\partial^2 V_P(\boldsymbol{x},\beta)}{\partial \beta^2}=\frac{6}{\beta^2}Q(\boldsymbol{x},\beta).
\end{equation}

When $T\to\infty$, the polymer collapses onto itself and resembles a classical particle. For this limiting case, a single replica can then be used again. On the contrary, when $T\to 0$, the spring interaction between replicas decreases and the polymer spreads over: the number of replicas needed also increases in order better to represent the quantum positional uncertainty \cite{tuckerman_statistical_2010}.

By analogy with the classical case (Eq. \eqref{eq_decomp_Z}), we denote by $Z_\textrm{k}^P$ the prefactor in Eq. \eqref{eq_quantum_partition_function}: 
\begin{equation}
    Z_\textrm{k}^P(\beta)=\left(\frac{P m}{2\pi\beta\hbar^2}\right)^{3NP/2}.
\end{equation}
The corresponding contributions to the internal energy and to the heat capacity are \cite{szekeres_direct_2018} $U_\textrm{k}^P(\beta)=\frac{3NP}{2\beta}$ and $C_\textrm{v,k}^P(\beta)=\frac{3NPk_\textrm{B}}{2}$. The position-dependent term is thus written
\begin{equation}\label{eq_zc_q_vp}
Z_\textrm{c}^P(\beta)=\idotsint d\boldsymbol{x}_1~...~d\boldsymbol{x}_P~\mathrm{e}^{-\beta V_P},    
\end{equation}
where $V_P$ is defined in Eq. \eqref{eq_potentiel_PI}. In general, $Z_\textrm{c}$ does not have an analytical expression. We therefore use the nested sampling algorithm to compute it, either with the direct or with the extended method. Note that there are two distinct convergence parameters: the number of live points $K$, as in the classical case, and the number of replicas $P$ that has to be large enough to correctly represent the NQEs.

In Sections \ref{sssec_dir_method_pres} and \ref{sssec_ext_method_pres}, we have seen that, due to the temperature dependence of the potential, the discrete expressions to recover the internal energy and the heat capacity are more complex than those from Eqs. \eqref{eq_u_nf_class} and \eqref{eq_cv_nf_class}, as they include new terms appearing whenever the effective potential depends on $\beta$. In Appendix \ref{app_PI_discrete_expressions}, we explicitly provide those expressions, in both direct and extended methods, in the particular case of the path-integral potential $V_P(\boldsymbol{x},\beta)$ (Eq. \eqref{eq_potentiel_PI}). Implementing the direct method is straightforward: nested sampling explorations are performed at the temperatures of interest and the thermodynamic properties are computed for each exploration. However, the implementation of the extended method is more complex and is discussed in the next section.
\subsection{Extended path-integral algorithm implementation}\label{ssec_extended_method}
\subsubsection{Choice of variables to explore the extended space}\label{sssec_extended_method_sampling}
\begin{figure}[h!]
    \centering
    \includegraphics[scale=1,trim=0 0 0 0,clip]{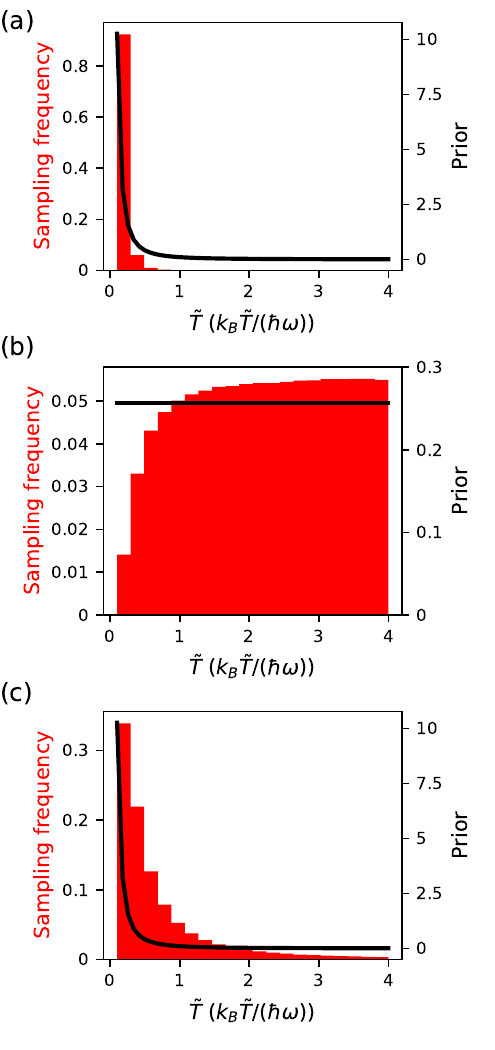}
    \caption{Sampling frequency for the harmonic potential with $N=1$ and $P=2$ (a) before the change of variables using $\tilde{\beta}$ as the explored parameter, (b) after the change of variables using $1/\tilde{\beta}$ as the explored parameter and (c) after the change of variables using $\tilde{\beta}$ as the explored parameter. We have $\tilde{T}=1/(k_\textrm{B}\tilde{\beta})$. The prior is indicated by the black curve.}
    \label{fig_hist_harm_2}
\end{figure}

Before using the extended method in practice, we present how to sample points using this method. Indeed, simply turning the auxiliary temperature $\tilde{\beta}=1/(k_\textrm{B} \tilde{T})$ into an additional parameter to explore in the nested sampling procedure with a uniform prior would result in an imbalanced sampling due to the particular form of the potential $Q$ in Eq. \eqref{eq_potentiel_PI}. This would cause the low auxiliary temperatures $\tilde{T}$ to be overrepresented in the sampling with respect to high $\tilde{T}$. This affects the efficiency of the method and would require extremely large numbers $K$ of live points to converge. An example is provided in Figure \ref{fig_hist_harm_2} (a) for the harmonic potential.
We see that the sampling of the temperature is imbalanced, with low temperatures being more sampled than high temperatures.

In order to overcome this problem, we change variable to obtain a more balanced sampling between low and high auxiliary temperatures. We denote
\begin{equation}
    \Tilde{\boldsymbol{y}}_i=\frac{\boldsymbol{x}_i-\Bar{\boldsymbol{x}}}{\lambda_P(\tilde{\beta})}~~~1\leq i \leq P,
\end{equation}
with $\Bar{\boldsymbol{x}}=\frac{1}{P}\sum_{i=1}^P \boldsymbol{x}_i$ the centroid of the replicas and 
\begin{equation}\label{eq_lambda_p}
    \lambda_P^2(\tilde{\beta})=\frac{\hbar^2\tilde{\beta}^2}{mP}.
\end{equation}
The variable $\Tilde{\boldsymbol{y}}_i$ is the position of the $i$-th replica with respect to the centroid, normalized by the factor $\lambda_P$, used to remove the temperature dependence of the harmonic interaction between replicas, as can be seen in the resulting expression for the potential
\begin{align}
V_P(\boldsymbol{x}_1,...,\boldsymbol{x}_P,\tilde{\beta})&=V_P(\bar{\boldsymbol{x}},\tilde{\boldsymbol{y}}_1,...,\tilde{\boldsymbol{y}}_{P},\tilde{\beta})\\
&=\frac{1}{2}\sum_{i=1}^P(\tilde{\boldsymbol{y}}_i-\tilde{\boldsymbol{y}}_{i+1})^2+\frac{1}{P}\sum_{i=1}^P V(\bar{\boldsymbol{x}}+\lambda_P(\tilde{\beta})\tilde{\boldsymbol{y}}_i).  \label{eq_potential_PI_transf}  
\end{align}
In this case, the position parameters are the position of the centroid $\Bar{\boldsymbol{x}}$ and of the first $P-1$ replicas $\{\tilde{\boldsymbol{y}}_i\}_{1\leq i<P}$. The $P$-th replica can be determined from the others as $\Tilde{\boldsymbol{y}}_P=-\sum_{i=1}^{P-1}\Tilde{\boldsymbol{y}}_i$. The auxiliary temperature dependence has been moved from the interaction between replicas $Q$ to the potential energy term $V$. Figure \ref{fig_hist_harm_2} (b)--(c) shows the sampling distribution after the transformation using $\tilde{\beta}$ and $1/\tilde{\beta}$ as the extra parameters with uniform prior. We see that the change of variable in the prior distribution results in a more balanced sampling.  Furthermore, the sampling frequency approaches the prior very closely when sampling with the uniform prior on $1/\tilde{\beta}$ (Figure \ref{fig_hist_harm_2} (c)). This shows that the choice of the function of $\tilde{\beta}$ to explore with uniform prior is also important. It is important to note that this change of variables and the choice of the parameter with uniform prior does not affect the fully converged results that would be obtained in the limit of an infinite number of live points $K$. However, an appropriate choice of variables can greatly reduce the number of live points needed in practice to converge and therefore the computational cost of the exploration.
\subsubsection{Choice of the upper and lower bounds of the parameters}
\begin{figure}
    \centering
    \includegraphics[width=3.3in]{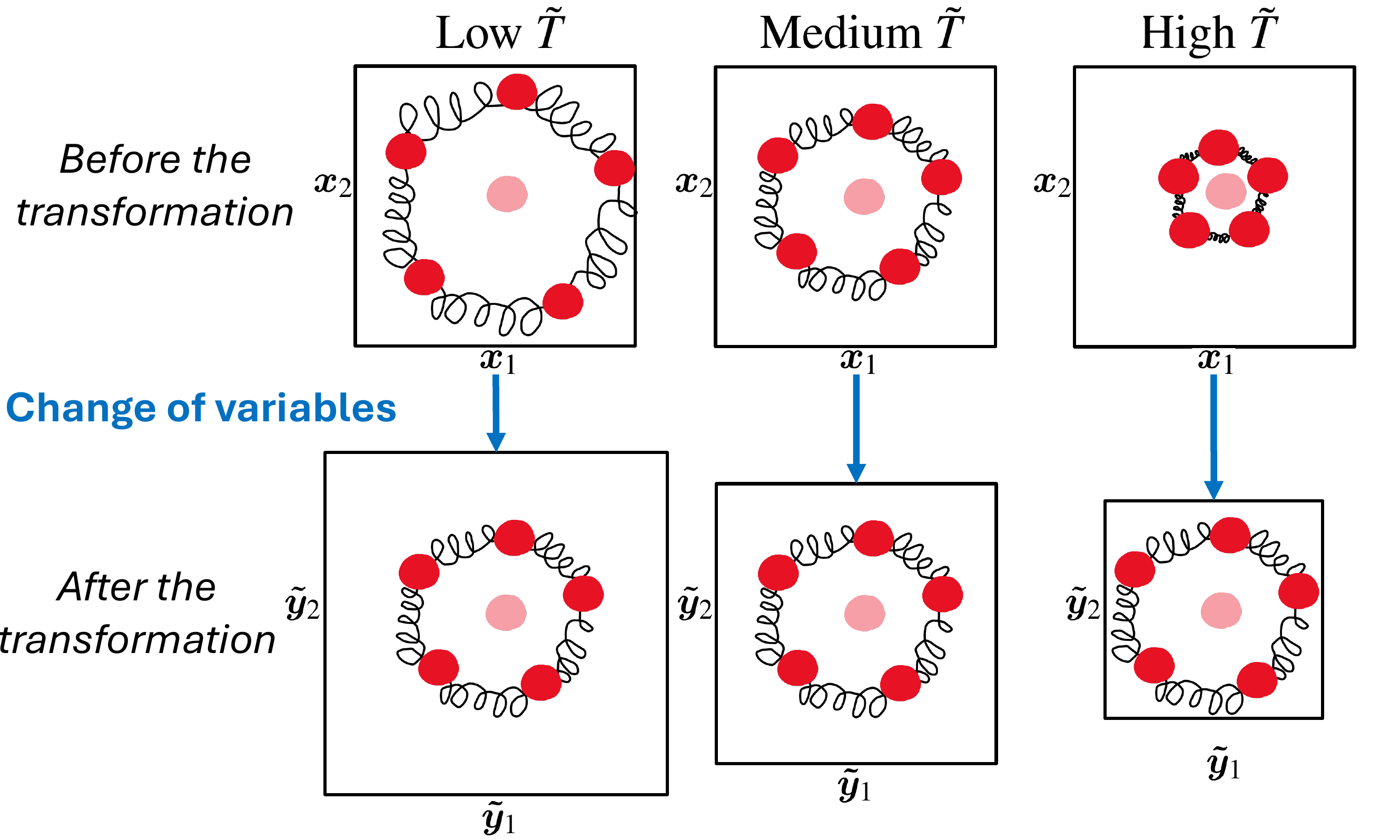}
    \caption{Representation of the parameter transformation given by Eqs \ref{eq_lambda_p}--\ref{eq_potential_PI_transf} (for $P=5$). Before the transformation, the polymer is in a box of fixed size (the same for all temperatures). In that case, the extension of the polymer depends on the temperature: the polymer contracts with increasing temperature. After the transformation, the extension of the polymer is the same for all temperatures (see first term of Eq. \ref{eq_potential_PI_transf}), however the size of the box is then proportional to $\lambda_P$, hence to the temperature.}
    \label{fig_rep_param_transf}
\end{figure}
With the change of variables presented above, the temperature $\tilde{\beta}$ affects the distance between the replicas and the centroid. \texttt{Nested\_fit} requires that we fix a pre-determined interval in which the parameters evolve: for high $\tilde{\beta}$ (low temperatures), the replicas can go further from the centroid than for low $\tilde{\beta}$ (high temperatures). Hence, choosing the lower and upper bounds of the parameters is not as straightforward as for the direct method. Indeed, without transformation, we simply take all replicas to be in an identical box of fixed size. With the transformation, the centroid is also placed in a box of fixed size. The difficulty arises when choosing the bounds for the $\Tilde{\boldsymbol{y}}_i$, that is, the position of the replicas relative to the centroid. Indeed, if $\Tilde{\boldsymbol{y}}_i$ is in a cubic box of side $a$ centered in 0, then the corresponding replica $\boldsymbol{x}_i$ is in a cubic box of size $a\lambda_P(\tilde{\beta}) $ centered in $\Bar{\boldsymbol{x}}$: the volume of space that is sampled by the replicas therefore depends on the temperature. The impact of the parameter transformation on the selection of the box size is represented in Figure \mbox{\ref{fig_rep_param_transf}}. We will indicate how to choose these bounds in practice for the two examples studied in this work: the quantum harmonic potential and Lennard-Jones clusters.

\section{Harmonic systems}\label{sec_harmonic}
For the quantum harmonic potential, the exact expressions for the internal energy and the heat capacity are known analytically\cite{tuckerman_statistical_2010, liboff_introductory_2003} (see Section \ref{ssec_harm_exact_expressions}), which makes it a good test for the implementation of nested sampling for studying NQEs.

\subsection{Exact equations}\label{ssec_harm_exact_expressions}
As a first application of the path-integral nested sampling method, we consider the following harmonic potential for one particle ($N=1$):
\begin{equation}
V(\boldsymbol{x}) = \frac{m\omega^2}{2} (x^2+y^2+z^2),
\end{equation}
with $\boldsymbol{x}=(x,y,z)$ the position vector of the particle. Consequently, the expression of $V_P$ in Eq. \eqref{eq_potentiel_PI} gives
\begin{align}
V_P(\boldsymbol{x}_1,...,\boldsymbol{x}_P,\beta)=&\frac{mP}{2\hbar^2\beta^2}\sum_{i=1}^P(\boldsymbol{x}_i-\boldsymbol{x}_{i+1})^2 \nonumber\\
&+\frac{1}{P}\sum_{i=1}^P \frac{m\omega^2}{2} (x_{i}^2+y_{i}^2+z_{i}^2),\label{eq_VP_harm}
\end{align}
where $m$ is the particle mass, $\omega$ the frequency and $\boldsymbol{x}_i=(x_i,y_i,z_i)$ is the three-dimensional position vector of the $i$-th replica. In practice, we use reduced units, i.e., we take $m$ and $\omega$ to be the units of mass and frequency. Hence, in reduced units, we have $m=1$ and $\omega=1$. We also take $\hbar=1$. Furthermore, we use the reduced temperature $\theta=k_\textrm{B}T/(\hbar\omega)$ and reduced dimensionless length units by normalizing positions by the factor $\sqrt{\hbar/(m\omega)}$. For the harmonic potential, the values of the lower and upper bounds of the parameters are not important as long as the sampling space is big enough not to be affected by finite size effects at the temperatures we are sampling (see Ref. \onlinecite{maillard_assessing_2023} for the study in the classical case). This is not the case for all potentials, as we will see when studying Lennard-Jones clusters: indeed, the harmonic potential is a confining potential, contrary to the Lennard-Jones potential which is null at infinite interparticle distance. Therefore, for the direct method, all replicas are placed in a finite box of size $L=30$ in reduced units; for the extended method, the centroid and the $\Tilde{\boldsymbol{y}}_i$ ($1\leq i\leq P-1$) are also confined in a box of side $L=30$ in reduced units.

In one dimension, the system can take the energy values \cite{tuckerman_statistical_2010} $E_n=(n+1/2)$  for $n=0,1,2,...$, corresponding to the eigenvalues of the Hamiltonian $H$. In the quantum setting, the partition function is given by\cite{tuckerman_statistical_2010} $Z(\beta)=\mathrm{Tr}\left(\exp(-\beta H)\right)$, hence
\begin{equation}\label{eq_Z_harm_Q}
    Z(\beta)=\sum_{n=0}^\infty \mathrm{e}^{-\beta E_n}=\frac{1}{2\sinh\left(\frac{1}{2\theta}\right)}.
\end{equation}
Consequently, we can compute the exact internal energy and heat capacity: for a three-dimensional system with one particle, the internal energy and heat capacity of the one-dimensional particle are multiplied by $3$, giving \cite{tuckerman_statistical_2010, liboff_introductory_2003}
\begin{equation}\label{eq_u_harm_q}
U(\beta)=\frac{3}{2}\coth\left(\frac{1}{2\theta}\right)
\end{equation}
and
\begin{equation}\label{eq_cv_harm_q}
C_v(\beta)=\frac{3}{4 \theta^2\sinh^2\left(\frac{1}{2\theta}\right)}k_\textrm{B},
\end{equation}
respectively.

Furthermore, for the harmonic potential, we can compute analytically the exact values of the heat capacity for a finite value of $P$ in particular,
\begin{equation}\label{eq_cv_harm_p_2}
C_v^{P=2}(\beta)=3k_\textrm{B}\left(2-\frac{4\theta^4+\frac{3}{4}\theta^2}{(2\theta^2+\frac{1}{8})^2}\right)
\end{equation}
and
\begin{equation}\label{eq_cv_harm_p_4}
C_v^{P=4}(\beta)=3k_\textrm{B}\left(4-2\frac{\theta^4+\frac{3}{32}\theta^2}{(\theta^2+\frac{1}{32})^2}-\frac{4\theta^4+\frac{3}{16}\theta^2}{(2\theta^2+\frac{1}{32})^2}\right)
\end{equation}    
for $P=2$ and $P=4$, respectively. The derivation of these expressions is given in Appendix \ref{app_harm_P_der}.

In contrast with the classical case, the quantum heat capacity in Eq. \eqref{eq_cv_harm_q} goes to zero in the low temperature limit. However, for finite numbers of replicas, the expressions \eqref{eq_cv_harm_p_2} and \eqref{eq_cv_harm_p_4} still tend to finite value for $T\to0$, with $C_v^{P=2}\to6k_\textrm{B}$ for $P=2$ and $C_v^{P=4}\to12k_\textrm{B}$ for $P=4$ for one particle in three dimensions. This corresponds to the classical heat capacity for a system with $6P$ degrees of freedom.

Before using \texttt{nested\_fit} with both the direct and extended methods on the harmonic system, we first provide a theoretical test of the effect of the use of a smeared delta-function $f(\beta-\tilde{\beta}; \alpha)$ in the extended method.

\subsection{Theoretical test of smearing functions}\label{sssec_ext_method_theo_harm}
\begin{figure}
    \centering
    \includegraphics[scale=1, trim= 0 12 0 0, clip]{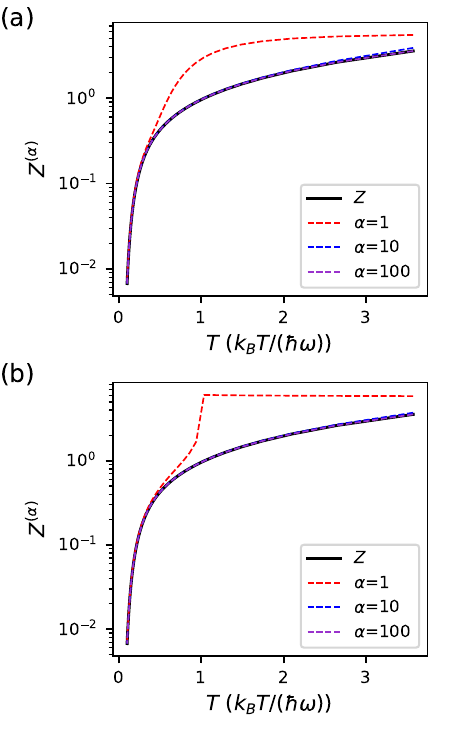}
    \caption{Comparison between $Z$ and the smeared-delta function approximation for (a) the Gaussian function (Eq. \eqref{eq_harm_gauss}) and (b) the rectangular function (Eq. \eqref{eq_harm_rect}). The curves corresponding to $Z$, $\alpha=10$ and $\alpha=100$ almost coincide at the graph scale.}
    \label{fig_ext_method_theo}
\end{figure}
In this section, we compute analytical results on the effect of the two smeared delta-functions $f(\beta-\tilde{\beta};\alpha)$ previously introduced. For that purpose, we use the quantum harmonic potential: for simplicity, we consider the one-dimensional case instead of the three-dimensional one. The exact partition function \cite{tuckerman_statistical_2010} is given in Eq. \eqref{eq_Z_harm_Q} and can be rewritten as an integral over $\tilde{\beta}$ as in Section \ref{sssec_ext_method_pres}:
\begin{equation}
    Z(\beta) = \int_0^\infty d\tilde{\beta} \delta(\beta-\tilde{\beta})Z(\tilde{\beta}).
\end{equation}
Replacing the delta-function by the Gaussian window gives the approximation
\begin{equation}\label{eq_harm_gauss}
    Z^{(\alpha)}(\beta) = \int_0^\infty d\tilde{\beta} \sqrt{\frac{\alpha}{\pi}} \mathrm{e}^{-\alpha (\beta-\tilde{\beta})^2} Z(\tilde{\beta}), 
\end{equation}
which depends on parameter $\alpha$. Replacing it by the rectangular window gives
\begin{equation}\label{eq_harm_rect}
    Z^{(\alpha)}(\beta) = \int_{\max(0, \beta-1/\alpha)}^{\beta+1/\alpha} d\tilde{\beta} \frac{\alpha}{2} Z(\tilde{\beta}).
\end{equation}

In the case of Eq. \eqref{eq_harm_gauss}, we compute $Z$ numerically. In the case of the rectangular window (Eq. \eqref{eq_harm_rect}), the integral yields
\begin{align}
    Z^{(\alpha)}(\beta) =& \frac{\alpha}{2\hbar\omega}\left[\log\left(\tanh\left(\frac{\hbar\omega(\beta+1/\alpha)}{4}\right)\right)\right.\nonumber\\
    &-\left.\log\left(\tanh\left(\frac{\hbar\omega\max(0, \beta-1/\alpha)}{4}\right)\right)\right].\label{eq_expl_harm_rect}
\end{align}

In order to compare the exact and the extended method results for the Gaussian window, we represent the curves for different values of $\alpha$ in Figure \ref{fig_ext_method_theo} (a). We take values of $k_\textrm{B} T=1/\beta$ between $0.2\hbar\omega$ ($\beta=5$) and $3.5\hbar\omega$ ($\beta=0.28$) which is the range of temperature studied in Section \ref{ssec_harm_dir}. In Figure \ref{fig_ext_method_theo}, we see that if we take $\alpha$ large enough ($\alpha \gtrsim 10$), $Z$ is correctly recovered for the range of temperature studied. For smaller values of $\alpha$, the curve is not correctly recovered for high temperatures. This is likely due to the fact that, for these combinations of $\alpha$ and $\beta$, the delta-function approximation gives significant weight to small values of $\tilde{\beta}$ where the integrand diverges. However, at high temperature a classical treatment of the system is possible (in that case at around $\sim 2.5 k_\textrm{B}T/(\hbar\omega)$, see Figure \ref{fig_harm_q_method_direct}), so that the behavior of $Z$ can be obtained by the simpler, classical version of the algorithm.

In Figure \ref{fig_ext_method_theo} (b), we represent the curves for different values of $\alpha$ for the same system, using the rectangular function. In that case, since $\log(\tanh(x))$ is not defined in 0, we replace $\max(0, \beta-1/\alpha)$ by $\max(10^{-5}, \beta-1/\alpha)$ in Eq. \eqref{eq_expl_harm_rect} to avoid numerical problems. This approximation can be done as, in practice, we sample between finite values of $\tilde{T}$ ($\tilde{\beta}>0)$. We observe a similar behavior with increasing $\alpha$ than what was observed for the Gaussian window. In that case too, we have a good recovery of $Z$ for $\alpha\gtrsim 10$.
\subsection{Direct method - comparison of \texttt{nested\_fit} with the exact cases} \label{ssec_harm_dir}
\begin{figure}
     \centering
     \includegraphics[width=3.3in,trim=10 13 0 0,clip]{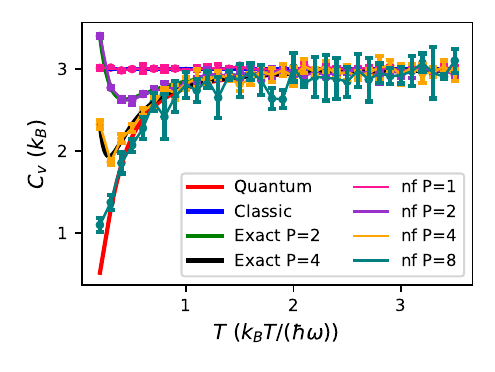}
     \caption{Heat capacity for the quantum harmonic potential with the direct method. The analytical results for the classical and quantum cases are plotted as full lines, as well as the formulas for two (Eq. \eqref{eq_cv_harm_p_2}) and four (Eq. \eqref{eq_cv_harm_p_4}) replicas ($P=2$ and $P=4$, respectively). Also shown are the numerical results obtained with \texttt{nested\_fit} (nf in the legend). The points correspond to the temperatures at which the heat capacities were computed.  Taking $\nu=\omega/(2\pi)=100$cm$^{-1}$, which is a typical vibration frequency in solids, the temperature range considered here would correspond to [290K,5000K]. The error bars were computed from four runs: for $P=1,2,4$, they are smaller than the symbols for most temperatures.}
     \label{fig_harm_q_method_direct}
\end{figure}

Figure \ref{fig_harm_q_method_direct} presents the results obtained for the heat capacity for $P=1,2,4,8$ with \texttt{nested\_fit} using the direct method (the exploration parameters are given in Appendix \ref{app_comp_details}, as will be the case for all applications in this work). The exact quantum, classical ($P=1$), $P=2$ and $P=4$ curves are also represented. As check, we perform one exploration per temperature for the $P=1$ curve with \texttt{nested\_fit}, even though this case is equivalent to the classical case and only one exploration can be performed for all temperatures. We can see that nested sampling is able to recover the exact curves.  We also see that when $P$ increases, the exact quantum curve is recovered down to lower temperatures, which is the expected behavior as the system is, so to say, "more quantum" at lower temperatures and thus requires more replicas to converge. Moreover, we see that at low temperatures, the curves for $P=2$ and $P=4$ (both exact and obtained via \texttt{nested\_fit}) start increasing, with the minimum of the curve being reached at a lower temperature for $P=4$ than for $P=2$. The unphysical increase in the heat capacity is the manifestation of the classical nature of the system for a finite number of replicas, whereas the limit $P\to\infty$ would be needed to simulate the quantum systems down to $T=0$. Therefore, for a finite number of replicas $P$, there is a threshold temperature $T_P^*$ such that $\partial C_\textrm{v}/\partial T < 0$ for $T<T_P^*$. Note that $T_P^*\to 0$ as $P\to\infty$. Finally, we observe that the nested sampling results fluctuate more and have a bigger statistical uncertainty for higher $P$, which is again expected as a higher value of $P$ means more degrees of freedom to explore. To reduce fluctuations, a larger number of live points $K$ is required (here, $K=1000$ for all $P$). The impact of $K$ on the recovery of the heat capacity will be studied in Section \ref{sec_LJ} for the more complex partition function of Lennard-Jones clusters, which has multiple minima. 

\subsection{Extended partition function method}
\subsubsection{Choosing the smeared delta function}\label{sssec_harm_q_choice_alpha}
We now perform simulations with \texttt{nested\_fit} using the extended method. For that, we use the path-integral harmonic potential with $P=2$. The computational details are given in Appendix \ref{app_comp_details}.
\begin{figure}
    \includegraphics[scale=1]{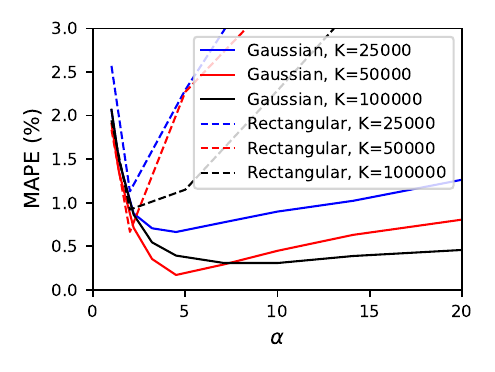}
    \caption{MAPE as a function of $\alpha$ for different values of $K$ with the Gaussian (full lines) and rectangular (dashed lines) windows.}
    \label{fig_harm_mape_alpha}
\end{figure}

\begin{figure*}
    \includegraphics[scale=1]{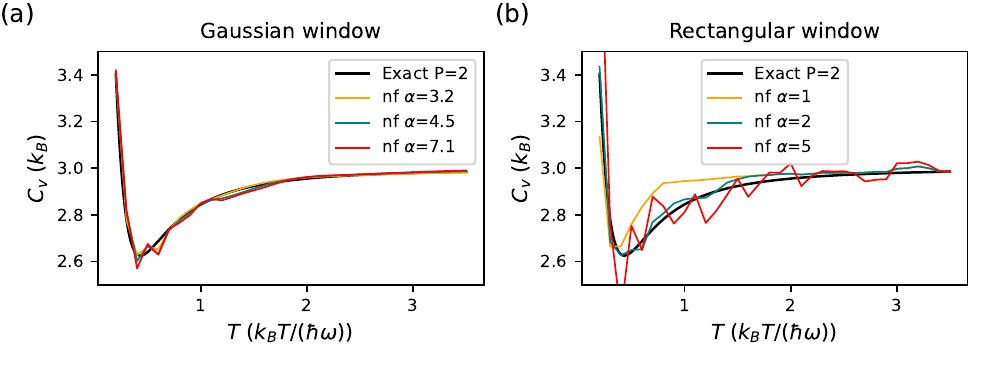}
    \caption{Comparison of the heat capacity obtained for the harmonic potential with the extended method for different values of $\alpha$ (a) for the Gaussian window and (b) the rectangular window. In each case only three values are represented for the readability of the graph.}
    \label{fig_harm_method_ext_comp_alpha}
\end{figure*}

For both approximations of the delta-function, we test different values of the parameter $\alpha$, corresponding to the size of the window where the inverse temperature $\tilde{\beta}$ is sampled. In order to quantify the computational accuracy, we adopt the Mean Absolute Percentage Error (MAPE) \cite{goodwin_asymmetry_1999}:
\begin{equation}\label{eq_MAPE}
\mbox{MAPE}=\frac{1}{n}\sum_{i=1}^n100\frac{|S_i-E_i|}{E_i},
\end{equation}
with $n$ the number of temperatures sampled, $E_i$ the expected results --- here the exact $P=2$ $C_\textrm{v}$ curve --- and $S_i$ the sampled points --- here the computed $C_\textrm{v}$ curve. In Figure \ref{fig_harm_mape_alpha}, we present the MAPE curves as a function of $\alpha$ for different values of $K$ for both Gaussian and rectangular windows. The $\alpha$ value for which the MAPE is the smallest changes with the window and number of live points $K$: for the Gaussian window, the smallest value of MAPE is obtained for $\alpha=4.5$ for $K=25000,50000$ and $\alpha=10$ for $K=100000$ while for the rectangular window, the smallest MAPE value is obtained for $\alpha\approx2$ for all $K$. In the following, we refer to this value of $\alpha$ as the "optimal" value, which is valid for a specific example and number of live points $K$. We point out that nested sampling provides a set of sampled points. These points are then post-processed by using distinct $g(\beta-\tilde{\beta};\alpha)$ windows: they are therefore issued from the same statistical set.

In Figure \ref{fig_harm_method_ext_comp_alpha}, we present the results obtained for the Gaussian and rectangular windows for $\alpha=3.2,4.5,7.1$ and $\alpha=1,2,5$ respectively, at $K=50000$, for which the smallest MAPE value is obtained. Again, all curves were obtained from the same exploration. In both cases, taking a small value (below their optimal one) of $\alpha$ does not correctly recover the heat capacity. Indeed, small values of $\alpha$ correspond to large temperature windows, which gives rise to a bias. For large $\alpha$ instead, we only select points that have an inverse temperature $\tilde{\beta}$ very close to $\beta$. Since the number of selected points decreases when $\alpha$ increases, large values of $\alpha$ result in a poor statistical sampling and large fluctuations (without bias). We see that the rectangular function fluctuates more than the Gaussian function. A possible reason for this behavior is that it either selects or rejects the sampled points, with the selected points all having equal weights. The rectangular window function therefore does not discriminate between different inverse temperatures $\tilde{\beta}$ depending on their distance from the target $\beta$, which is the case of the Gaussian function. Hence, from now on, we choose the Gaussian window for comparing the extended and direct methods.

\begin{figure}
    \centering
    \includegraphics[width=3.3in]{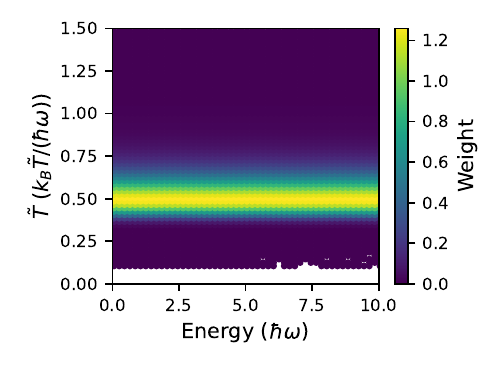}
    \caption{Plot of the auxiliary temperature $\tilde{T}$ against the energy of the sampled configurations for the $P=2$ and $K=50000$ case. The points are colored according to the weight given by the Gaussian approximation with $\alpha=5$ and $T=0.5(k_\mathrm{B}T)/(\hbar\omega)$.}
    \label{fig_sampling_aux_temp}
\end{figure}

In Figure \mbox{\ref{fig_sampling_aux_temp}}, we represent the auxiliary temperature against the energy of the sampled configurations, colored according to the weight given by the Gaussian approximation. We see that only a small proportion of points contribute significantly to the computation of the thermodynamic properties at a given temperature: this explains why we need such a huge number of live points to compute the thermodynamic properties at all temperatures with the extended method. Note that we have only represented the points for $\tilde{T}\leq 1.5 (k_\mathrm{B}\tilde{T})/(\hbar\omega)$ but we have sampled up to $\tilde{T}=4 (k_\mathrm{B}\tilde{T})/(\hbar\omega)$.

\subsubsection{Comparing with the direct method}\label{sssec_harm_q_comp_1_2}

\begin{figure}
    \centering
    \includegraphics[scale=1,trim=0 12 0 10,clip]{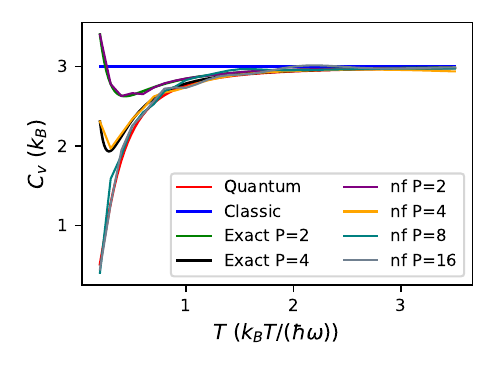}
    \caption{Heat capacity for the quantum harmonic potential obtained using \texttt{nested\_fit} (nf) with the extended method with $\alpha=4.5$ for $P=2$ and $\alpha=7.1$ for $P=4,8,16$. We used $K=50000$, $K=100000$, $K=100000$ and $K=200000$ for $P=2$, $P=4$, $P=8$ and $P=16$, respectively.}
    \label{fig_harm_q_method_ext}
\end{figure}

We now look at the performance of the extended method on the harmonic potential and compare our results with that obtained with the direct method (Figure \ref{fig_harm_q_method_direct}). We again look at the curves obtained with \texttt{nested\_fit} for $P=2,4,8$ as well as for $P=16$. We put a uniform prior on $\tilde{T}$ for $P=2,8,16$ and a uniform prior on $\tilde{\beta}$ for $P=4$. For the latter, the minimum of the $P=4$ exact curve (around $0.3k_\textrm{B}T/(\hbar\omega)$) was not correctly recovered when using a uniform prior on $\tilde{T}$. This shows that the choice of the prior, which, contrary to $\alpha$, has to be chosen before the exploration, has a noticeable impact on the resulting curves as was illustrated in Section \ref{sssec_extended_method_sampling}. The results are presented in Figure \ref{fig_harm_q_method_ext}. We see that, with the extended method, \texttt{nested\_fit} is able to recover the exact curves very well for $P=2,4$; for $P=8,16$ (for which we do not have reference curves), nested sampling converges to the exact quantum results (as expected for $P\to\infty$). The corresponding sampling frequency are given in the supplementary material (Figure 1).

\begin{figure}
    \centering
    \includegraphics[width=3.3in, trim=12 10 30 30, clip]{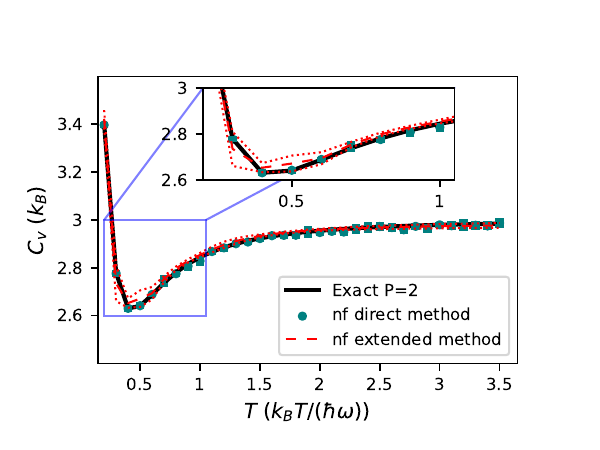}
    \caption{Comparison of the direct and extended methods with $\alpha=7.1$ on the quantum harmonic potential with $P=2$. The points correspond to the temperatures at which the heat capacities were computed with the direct method. The dotted red lines correspond to the first standard deviation for the extended run. The statistical uncertainties were computed from four runs.}
    \label{fig_harm_q_comp_direct_ext}
\end{figure}

Finally, we compare the computational performance of the extended and direct methods on the $P=2$ case. We present the results in Figure \ref{fig_harm_q_comp_direct_ext} using $K=25000$ live points for the direct method (one exploration per temperature) and $K=50000$ for the extended one (one exploration for all temperatures) with $\alpha=7.1$. Those choices result in similar MAPE values: 0.30\% for the direct method and 0.29\% for the extended method. Both methods correctly recover the exact curve and have error of the same order at high temperatures. At lower temperatures ($T<=0.6 (k_\mathrm{B}T)/(\hbar\omega)$), the extended method has larger error than the direct method by around a factor 10: we have less samples at those temperatures (see supplementary material, Figure 1 (a)). To obtain the heat capacity curves for 34 temperatures ranging from $k_\textrm{B}T=0.2\frac{k_\textrm{B}T}{\hbar\omega}$ to $k_\textrm{B}T=3.5\frac{k_\textrm{B}T}{\hbar\omega}$, the direct method requires $2.9\times10^9$ energy evaluations (i.e., approximately $8\times10^7$ energy evaluations per exploration) and the extended method requires $1.8\times10^8$ energy evaluations. The extended method is thus approximately eight times faster than the direct method. We observe that, even though we use more live points, the extended method requires much less evaluations --- by around a factor ten --- than the direct one. Indeed, in the former case, we only need a single exploration to compute the heat capacity at all temperatures while we need multiple explorations in the latter case. Furthermore, if we want to add a point to the curve within the studied temperature range, a new exploration must be done for the latter while for the former the heat capacity for the new temperature can be computed from the previously sampled points as the choice of a specific temperature window is a post-processing step. Note that the two energy functions are slightly different as, in the extended case, there is a change of variables which is not used with the direct method. This change of variables results in a small increase in computational cost for the function, that becomes negligible when the number of parameters increases or the potential is more complex than the harmonic potential. 

In conclusion, in this section, we have tested the extended partition function method on the harmonic potential. Mainly, we have looked at the impact of the choice of the smeared delta-function and of the corresponding parameter $\alpha$: we have seen that the Gaussian window leads to reduced fluctuations with respect to the rectangular one and that the optimal value of $\alpha$ depends on the number of live points used and the system studied. Furthermore, comparing with the direct method, we have seen that the extended method requires less energy evaluations, roughly by an order of magnitude. Finally, we have seen that the choice of the window and $\alpha$ is a post-processing step, contrary to the choice of the prior and number of live points $K$. This means that different $\alpha$ can be tested on one exploration but that a new exploration must be performed to test a new prior (more details on the choice of the prior will be given in Section \ref{ssec_LJ_13}). In the next section, we study the impact of $K$ on the recovery of the curves with both the direct and extended methods on the Lennard-Jones clusters that display a very large number of local energy minima.
\section{Lennard-Jones clusters}\label{sec_LJ}
\subsection{The potential model}
Lennard-Jones clusters are formed by $N$ particles interacting via
\begin{equation}
V(\boldsymbol{x}_i)=\sum_{\substack{1\leq k < j \leq N\\r_{i,kj}<r_c}} \left[V_\textrm{LJ}(r_{i,kj}) - V_\textrm{LJ}(r_\textrm{c})\right]
\end{equation}
with $r_\textrm{c}$ the cutoff radius, which removes interaction at infinite range, $r_{i,kj}$ the distance between the $i$-th replica of the $k$-th and $j$-th atoms and
\begin{equation}\label{eq_LJ}
V_\textrm{LJ}(r)=4\epsilon\left(\left(\frac{r_0}{r}\right)^{12}-\left(\frac{r_0}{r}\right)^{6}\right).
\end{equation}
We have that $r_0$ and $\epsilon$ are the parameters of the potential that depend on the considered atomic species. We take $r_\textrm{c}=3r_0$.  Furthermore, we use periodic boundary conditions in a large cubic simulation box of size $L\gg r_0$.

We adopt reduced units: the temperature is in units of $\frac{k_\textrm{B} T}{\epsilon}$ with $\epsilon=1$. To fix the size of the box, we consider the value of the density $\tilde{\rho}$ in unit of $r_0^{-3}$: $\tilde{\rho}=\frac{N}{(L/r_0)^3}r_0^{-3}$. We take $L=6$ and use $r_0$ to tune the density to a given desired value. The density thus characterizes the size of the space the particles evolve in, relative to the number of atoms $N$, which is fixed. In this work, $L$ is chosen so that the minimum image convention applies.

In the classical case, all Lennard-Jones systems are equivalent up to a rescaling of energies and distances, thus the results in reduced units are valid for all atomic species\cite{maillard_nested_2025} by inserting the actual value for $\epsilon$ and $r_0$. This is no longer true for the quantum case as changing the atomic species (and in particular the nuclei mass) changes the importance of the NQEs. This can be captured by the de Boer parameter \cite{de_boer_contribution_1938, de_boer_quantum_1948}:
\begin{equation}\label{eq_de_boer}
     \Lambda=\frac{\lambda_B}{r_0}.
\end{equation}
The parameter $\Lambda$ is the ratio between an effective the Broglie wavelength\cite{de_boer_contribution_1938, sevryuk_why_2010} $\lambda_B=\frac{h}{\sqrt{m\epsilon}}$, for particles of mass $m$ and typical energy $\epsilon$, and a characteristic length $r_0$ (of the order of the interparticle distance). Hence, a high value of parameter $\Lambda$ ($\Lambda \gtrsim1$) indicates that the quantum effects are important while a small value ($\Lambda \ll1$) indicates that they are less important and that a classical treatment of the system may be suitable. As we can see from Eq. \eqref{eq_de_boer}, the de Boer parameter depends on the mass. Therefore, in the following, for each example, we will fix $\hbar=1$ ($h=2\pi$) as well as the values of $\epsilon$ and $r_0$ and change the mass to have the de Boer parameter of the studied atomic species.

\subsection{7-atom Krypton cluster Kr$_7$}
\begin{figure}
    \centering
     \includegraphics[scale=1,trim=0 12 0 10,clip]{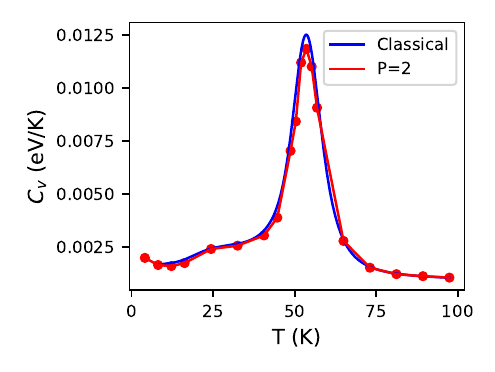}
     \caption{Heat capacity for the 7-atom Krypton cluster with the direct method. The points correspond to the temperatures at which the $P=2$ heat capacity was computed.}
    \label{fig_cv_lj_7_kr}
\end{figure}

We first consider a 7-atom Krypton cluster, which has a de Boer parameter of\cite{ashcroft_solid_1976} $\Lambda=0.10$. In that case, we take $\epsilon=1$, $r_0=0.659$ --- this corresponds to the classical case studied in Ref. \onlinecite{maillard_nested_2025} --- and a box of size $L=6~(\approx9.1r_0)$ for all replicas. From Eq. \eqref{eq_de_boer}, this fixes $m=h^2/(\Lambda^2r_0^2\epsilon)\approx 9091$ in reduced units. In Figure \ref{fig_cv_lj_7_kr}, we represent the classical and path-integral heat capacities as obtained with two replicas in the direct method. We see that the two curves almost coincide, within the statistical error. Hence, in this case, the classical treatment of the atoms is possible as using two replicas gives sensibly the same results as using one. This behavior is in accordance with the small de Boer parameter of Krypton.

The main peak in the graph at $T\sim50$K corresponds to the vapor-liquid transition, while the shoulder at $T\sim25$K can be associated to a liquid-like to solid-like state \cite{partay_efficient_2010}. The fact that $C_v$ starts to increase again at low $T$ for the path-integral curve is related to the finite value of $P$, as noted ahead in the harmonic oscillator case. 

\subsection{3-atom Neon cluster Ne$_3$}
We then consider the case of a 3-atom Neon cluster, for which the de Boer parameter is\cite{ashcroft_solid_1976} $\Lambda=0.59$. We take $\epsilon=1$, $r_0=1$ and a box of size $L=6r_0$ for all replicas. Since $\Lambda=0.59$, we take $m=113$. This system is used to study the convergence of the heat capacity with the number of replicas and with the number of live points, first with the direct method and then with the extended method. 
\subsubsection{Direct method}\label{par_ne_3_direct}
\begin{figure}[!b]
    \centering
    \includegraphics[scale=1,trim=0 15 0 0,clip]{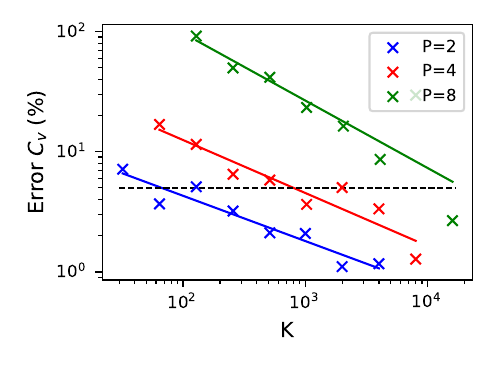}
    \caption{MAPE error for the 3-atom Neon cluster with $P=2,4,8$ with the direct method. The black dashed line represents an error of 5\%. The full lines are the linear log-log fits for each value of $P$. The slopes are $-0.377$, $-0.441$ and $-0.561$ for $P=2$, $P=4$ and $P=8$, respectively.}
    \label{fig_MAPE_error_method_1}
\end{figure}

First, for each $P$, we study the convergence with the number of live points $K$, starting by\footnote{This is done as \texttt{nested\_fit} may encounter problem with the Cholesky transformation of the covariance matrix otherwise.} $K>3NP$ ($3NP$ is the number of degrees of freedom). We then double this value until we obtain two consecutive heat capacity curves that look similar, that is, between which the MAPE (using the explorations with higher $K$ as reference) is less than 1.25\% for $P=2$, 1.5\% for $P=4$ and 3\% for $P=8$. The highest $K$ value considered is $K=8000$ for $P=2$, $K=16000$ for $P=4$ and $K=32000$ for $P=8$. Because for this system the exact analytical result is unknown, we consider the curve with the most live points and compare it to the curves obtained for smaller $K$. Again, we use the MAPE, with the curve obtained with the highest $K$ as reference. We set a threshold of 5\% for considering a curve to be converged. The MAPE is represented in Figure \ref{fig_MAPE_error_method_1} as a function of $K$ and for each value of $P$. 

We can see that, for the same value of $K$, the error is higher for higher $P$, as expected, as higher $P$ results in a higher number of degrees of freedom. In each case, we take as the first converged results the first point that is below the 5\% threshold (dashed line): $K=64$ for $P=2$, $K=1024$ for $P=4$ and $K=16000$ for $P=8$. We denote by $K_5$ the number of live points needed for reducing the MAPE below the 5\% threshold for $P$ replicas. The curves obtained for various $K$ values for $P=2,4,8$ are given in the supplementary material (Figure 2).  Finally, we expect the statistical uncertainty to scale as $1/\sqrt(K)$ \mbox{\cite{skilling_nested_2009, chopin_properties_2010}} and we indeed see that the MAPE decreases with the value of K with a slope of $\sim -0.5$ in log-log scale.

\begin{figure}
    \centering
    \includegraphics[scale=1]{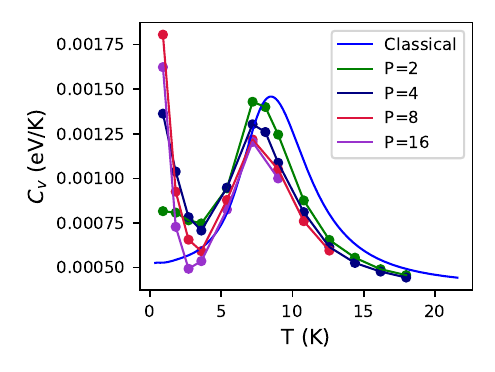}
    \caption{Heat capacity for the 3-atoms Neon cluster. The explorations with $K=8000$, $K=16000$, $K=32000$ and $K=64000$ are represented for $P=2$, $P=4$, $P=8$ and $P=16$, respectively.}
    \label{fig_cv_lj_3_ne_conv_P_16}
\end{figure}

The converged heat capacities for each $P$ are represented in Figure \ref{fig_cv_lj_3_ne_conv_P_16}. For $P=2,4,8$, we take the curves obtained with the highest $K$. For $P=16$, we use a large value of $K$ such that the curve coincides with the curve for $P=8$ at high temperatures. We observe a shift of the transition peak, corresponding to sublimation \cite{partay_efficient_2010}, towards lower temperatures when using $P>1$ replicas: it goes from being at around 10K for $P=1$ to around 7K for $P>1$. An analogous shift towards lower temperatures is due to NQEs and has been observed in Ref. \onlinecite{frantsuzov_quantum_2004} for the 13-atom Neon cluster. Furthermore, as for the harmonic potential, we see that when increasing $P$, the curves tend to coincide at a lower temperature for two consecutive values of the number of replicas. As for the harmonic oscillator, there is an increase of the heat capacity at very low $T$ (we suspect that this increase is caused by using a finite value of $P$). Indeed, this behavior appears at lower temperature for increasing $P$, as was the case for the harmonic potential.

To summarize, a system for which NQEs are more important --- high de Boer parameter, low temperature --- requires a larger number of replicas to recover the heat capacity curves. We have also seen that the increase of the number of replicas $P$ leads to an increase of the number of live points $K$ resulting in an increase of the computational cost of the exploration. As this procedure must be repeated at each temperature within the direct method, the total amount of computer time can be considerable. In the next section, we will show how it could be reduced by sampling the extended partition function in a single exploration for all temperatures.
\subsubsection{Extended method}\label{sssec_ne_3_method_2}

\begin{figure}
    \centering
    \includegraphics[scale=1,trim=0 12 0 10,clip]{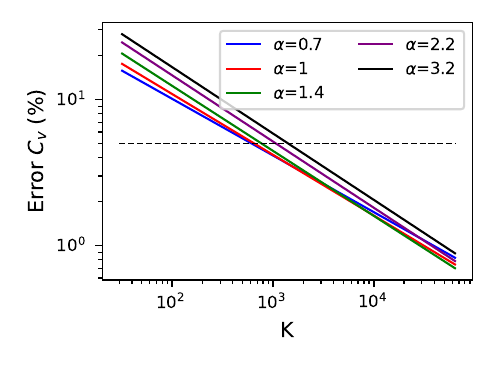}
    \caption{MAPE error for the 3-atoms Neon cluster with $P=2$ for different values of $\alpha$. The black dashed line represents an error of 5\%. The full lines are the linear log-log fits for each value of $P$. We only show the fits obtained from the points for readability. The slopes are $-0.388$, $-0.416$, $-0.445$, $-0.453$ and $-0.454$ for $\alpha=0.7$, $\alpha=1$, $\alpha=1.4$, $\alpha=2.2$ and $\alpha=3.2$, respectively.}
    \label{fig_MAPE_LJ_3_2_alpha}
\end{figure}

The curves obtained for the highest number of live points ($K=8000$ for $P=2$, $K=16000$ for $P=4$ and $K=32000$ for $P=8$) with the direct method are considered as a reference for the heat capacity as obtained with the extended method. Here, we do not consider the two lowest temperatures studied using the direct method as they exhibit this increase of the heat capacity (likely due to using a finite value of $P$) and are not converged (in terms of $P$). Moreover, we only adopt the Gaussian window and not the rectangular one as we have seen that the latter leads to larger fluctuations than the former. Furthermore, we use the change of variables presented in Section \ref{ssec_extended_method}. In that case, choosing the bounds of the parameters is more complex. For Lennard-Jones clusters, the size of the box affects the density and hence the heat capacity curves: the choice of the bounds is discussed in detail in Appendix \ref{app_box_size}.

\begin{figure}
    \includegraphics[scale=1,trim=0 10 0 0,clip]{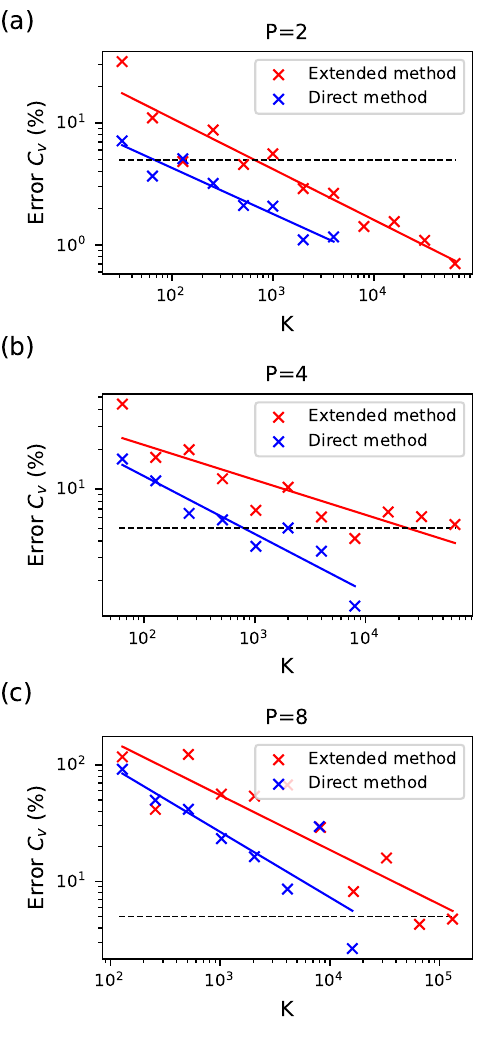}
    \caption{MAPE error for the 3-atoms Neon cluster with $P=2,4,8$ with the direct and extended methods. The black dashed line represents an error of 5\%. The full lines are the linear log-log fits. We have $\alpha=1$ for $P=2,4$ and $\alpha=1.4$ for $P=8$. For the extended method, the slopes of the fits are $-0.416$, $-0.267$ and $-0.468$ for $P=2$, $P=4$ and $P=8$, respectively. For the direct method, the slopes are indicated in Figure \ref{fig_MAPE_error_method_1}.}
    \label{fig_MAPE_LJ_3_P}
\end{figure}
For the harmonic potential, the optimal value of $\alpha$ depends on the example studied and number of live points $K$. Therefore, first we look at the evolution of the MAPE with $K$ for different values of $\alpha$ for $P=2$ (Figure \ref{fig_MAPE_LJ_3_2_alpha}), starting with the same number of live points considered for the direct method, which we then double in order to check the convergence. We find that, for this range of $K$ values, $\alpha=0.7,1,1.4$ give similar errors, which are smaller than for higher values of $\alpha$. Similar curves for $P=4,8$ are provided in the supplementary material (Figure 3) Therefore, in the following, we use $\alpha=1$ for $P=2,4$ and $\alpha=1.4$ for $P\geq8$.

\begin{figure}
    \includegraphics[scale=1]{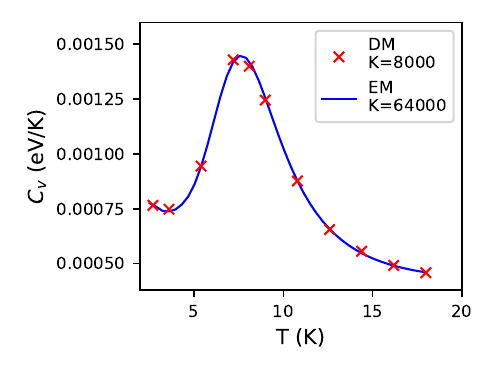}
    \caption{Comparison of the heat capacity for the 3-atoms Neon cluster between the direct method (DM) and the extended method (EM) for $P=2$. We have $\alpha=1$.}
    \label{fig_cv_lj_3_ne_conv_method_ext}
\end{figure}

\begin{figure}
    \centering
    \includegraphics[scale=1]{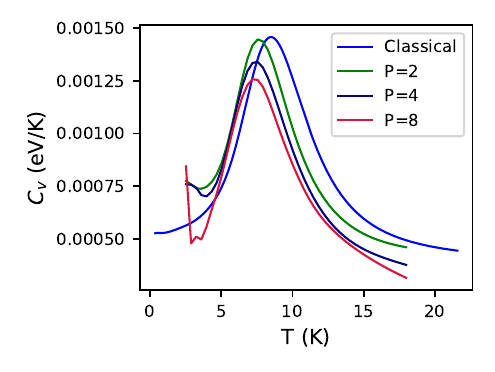}
    \caption{Heat capacity for the 3-atoms Neon cluster using the extended method. The explorations with $K=64000$, $K=64000$ and $K=131072$ are represented for $P=2$, $P=4$ and $P=8$, respectively.}
    \label{fig_cv_lj_3_ne_conv_P_ext}
\end{figure}

Secondly, we look at the MAPE for $P=2,4,8$ to find the minimum number of live points required for convergence. We represent the error in Figure \ref{fig_MAPE_LJ_3_P} for $P=2,4,8$ to compare the results obtained when using the direct and extended methods. First, we can see that, with the extended method, we need a larger $K$ to go below the 5\% error threshold, which is expected as we sample all temperatures at once. We therefore need more points to sample the energy surface at all temperatures. For the converged value of $K$, we again take the first point below the 5\% line, that is, $K=512$ for $P=2$, $K=8000$ for $P=4$ and $K=65536$ for $P=8$. We can see that we need 4 to 8 times more live points than with the direct method (for which the values were $K=64$, $K=1000$ and $K=16000$, respectively). In Figure \ref{fig_cv_lj_3_ne_conv_method_ext}, we see that the extended method is able to recover the heat capacity curve, as found with the direct method. More details are given in the supplementary material (Figure 4). The two methods therefore converge towards the same result. In Figure \ref{fig_cv_lj_3_ne_conv_P_ext}, the converged heat capacities for each $P$ are represented as obtained with the extended method using the highest number of live points, similarly to Figure \ref{fig_cv_lj_3_ne_conv_P_16} for the direct method. Furthermore, we see that, as for the direct method, the MAPE decreases with the value of K with a slope of $\sim -0.5$ in log-log scale, as expected.

Finally, we look at the number of energy evaluations made in the direct and extended methods for $P=2,4,8$ for the converged explorations. The results are shown in Table \ref{tab_tps_LJ_Q_methods}. We can see that to obtain a converged curve, the extended method requires fewer energy evaluations, which is expected as only one exploration is needed. Furthermore, with the direct method, we only access the sampled temperature while the extended method provides the thermodynamic properties for all temperature within the studied range. Hence, if we want to compute the thermodynamic properties at a new temperature, it is simply a post-processing step for the extended method while a new exploration must be performed for the direct method.

On this example, we have seen that the extended method requires more live points than the direct method. We have also seen that the optimal value of $\alpha$ depends on the example studied.

\begin{table}
    \caption{Total number of energy evaluations made by \texttt{nested\_fit} for the 3-atom Neon cluster for both methods in the converged cases (according to MAPE with a threshold of 5\%). In the direct case, 13 explorations were performed for $P=2,4$ and 9 for $P=8$. Twenty cores were used for $P=2,4$ and 64 for $P=8$ (see main text).}
    \label{tab_tps_LJ_Q_methods}
    \begin{ruledtabular}
    \begin{tabular}{lcc}
      Number of replicas~~~~& Direct method~~~~& Extended method \\
      \hline
      $P=2$  & $2.6\times 10^7$ & $1.4\times 10^7$ \\
      $P=4$  & $1.7\times 10^9$ & $7.7\times 10^8$ \\
      $P=8$  & $5.8\times 10^{10}$ & $3.8\times 10^{10}$ \\
    \end{tabular}    
    \end{ruledtabular}
\end{table}
\subsection{13-atom Neon cluster Ne$_{13}$}\label{ssec_LJ_13}
We now turn to the 13-atom Neon cluster Ne$_{13}$. This example was previously studied in other works such as Refs. \onlinecite{frantsuzov_quantum_2004, predescu_heat_2003, calvo_quantum_2001} using different sampling schemes (Gaussian wave-packet Monte Carlo \cite{frantsuzov_quantum_2004}, path-integral Monte-Carlo \cite{predescu_heat_2003} and the use of effective potentials and the harmonic superposition approximation \cite{calvo_quantum_2001}). We therefore compare our results using nested sampling with those obtained by Refs. \onlinecite{frantsuzov_quantum_2004, predescu_heat_2003}: indeed, both use the same implementation of the Lennard-Jones potential.
\begin{figure*}
    \centering
    \includegraphics[scale=1,trim=0 10 0 0,clip]{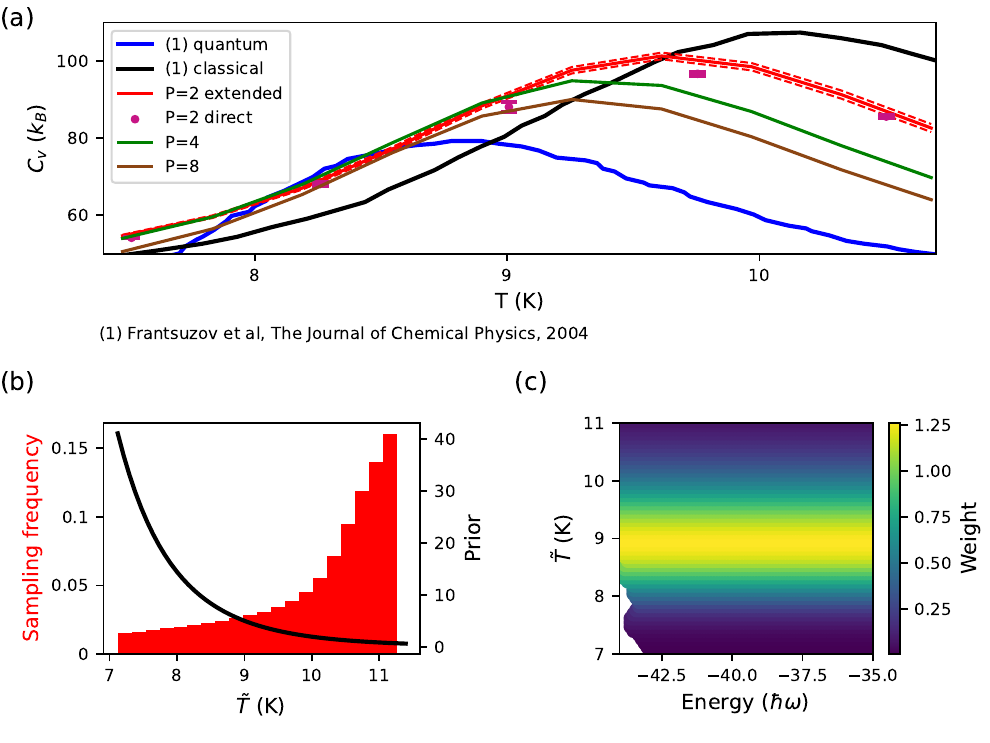}
    \caption{(a) Heat capacity for the 13-atoms Neon cluster. Comparison with the curve FC-VGW-MC in Ref. \onlinecite{frantsuzov_quantum_2004} using $\tilde{\beta}^8$ as an extra parameter. The curves for $P=2$ and $P=4$ are converged in $K$. The dashed red lines correspond to the first standard deviation for the $P=2$ extended run. For the direct runs, the error bars are smaller than the symbols. The statistical uncertainties were computed from four runs. (b) Sampling frequency of the temperature for the 13-atom Neon cluster as yielded by the extended method using $\tilde{\beta}^8$ as an extra parameter for $P=4$. We have $\tilde{T}=1/(k_\textrm{B}\tilde{\beta)}$. The prior $\propto \tilde{T}^{-9}$ is indicated by the black curve. (c) Plot of the auxiliary temperature $\tilde{T}$ against the energy of the sampled configurations for the $P = 2$ case using $\tilde{\beta}^8$ as an extra parameter. The points are colored according to the weight given by the Gaussian approximation with $\alpha = 5$ and $T = 9$ K.}
    \label{fig_lj_13_q_method_ext_beta8}
\end{figure*}

In this case, the potential $V$ in $V_P$ (Eq. \eqref{eq_potentiel_PI}) is the potential presented in Ref. \onlinecite{predescu_heat_2003}, also used in Ref. \onlinecite{frantsuzov_quantum_2004}:
\begin{equation}\label{eq_conf_LJ_potential}
    V(\boldsymbol{x})=V_{\textrm{LJ}}(\boldsymbol{x})+\sum_{i=1}^N V_{\textrm{c}}(\boldsymbol{x}_i),
\end{equation}
where $V_{\textrm{LJ}}$ is the full Lennard-Jones potential from Eq. \eqref{eq_LJ}, $\boldsymbol{x}$ the position vector of all $N$ atoms, $\boldsymbol{x}_i$ the position vector of atom $i$ and $V_\textrm{c}$ a confining potential which has the following form
\begin{equation}
    V_\textrm{c}(\boldsymbol{x}_i)=\epsilon\left(\frac{|\boldsymbol{x}_i-\boldsymbol{x}_\textrm{c}|}{2r_0}\right)^{20}
\end{equation}
with $r_0$ and $\epsilon$ the Lennard-Jones parameters and $\boldsymbol{x}_\textrm{c}=\frac{1}{N}\sum_{i=1}^N\boldsymbol{x}_i$ the center of mass of the cluster. The confining potential is used to prevent the atoms from leaving the cluster \cite{predescu_heat_2003} (in Ref. \onlinecite{frantsuzov_quantum_2004}, the confining potential is replaced by the constraint that the initial positions have to be within a distance of $2r_0$ from the center of mass). We sample the extended partition function with $\alpha=1$. In that case, it is easier to choose the bounds of the parameters than for Lennard-Jones clusters without confining potential: we simply need the sampling space to be large enough so that the confining potential prevents the atoms from leaving the space. Hence, we take $\Bar{\boldsymbol{x}}$ and $\tilde{y}_i$ for $i=1,..,P-1$ in a box of side $6r_0$ centered in 0: we can therefore have $|\boldsymbol{x}_i-\boldsymbol{x}_\textrm{c}|=6r_0$ which gives $V_\textrm{c}(\boldsymbol{x}_i)\approx3\times10^9\epsilon$. The classical curve, both from \texttt{nested\_fit} and Ref. \onlinecite{predescu_heat_2003} are given in the supplementary material (Figure 5)

We compare our results with the FC-VGW-MC (fully coupled variational Gaussian wave-packet Monte Carlo) method in Ref. \onlinecite{frantsuzov_quantum_2004} that was close to the results found in Ref. \onlinecite{predescu_heat_2003} --- in both cases, the error on the curve is around 1--2$k_\textrm{B}$. 
In this case, we put a uniform prior on $\tilde{\beta}^8$. As discussed in Section \ref{sssec_extended_method_sampling}, the choice of the prior for $\tilde{\beta}$ has an impact on the sampling distribution of the temperature. There we tested uniform priors on $\tilde{\beta}^{2s}$. For $s<4$, low temperatures are poorly sampled (we show the sampling distribution and resulting heat capacity for $s=0.5$ in Figure 6 in the supplementary material). For $s\geq 4$, low temperatures are sufficiently sampled and no apparent bias appears in the statistical quantities (Figure \ref{fig_lj_13_q_method_ext_beta8} (b)). In Figure \mbox{\ref{fig_lj_13_q_method_ext_beta8}} (c), we see that, when computing the thermodynamic properties at 9 K, there is only a small area of significant Gaussian weight with no sampled points (corresponding to points with very low energies and low temperatures). This is an improvement to the sampling obtained using a uniform prior on $\tilde{\beta}$ where we have few points with an auxiliary temperature below 8.5 K at low energies (supplementary material, Figure 6 (c)). Looking to the heat capacity curves (Figure \ref{fig_lj_13_q_method_ext_beta8} (a)), we see that, with increasing $P$, the curves are closer to the quantum one and farther from the classical one. However, the curve of Ref. \onlinecite{frantsuzov_quantum_2004} is not fully recovered by the extended method, even with $P=8$. Higher values of $P$ are therefore needed. When $P$ increases, the sampling frequency becomes more unbalanced towards high temperatures. Possibly, a uniform prior over $\tilde{\beta}^{2s}$ ($s>4$) will be necessary to sample the space correctly for higher $P$. Finally, the statistical uncertainties obtained for the direct and extended methods are of the same order (the extended method uses double the number of live points compared to the direct method).

With this example, we have seen that the prior we give to the auxiliary temperature greatly impacts the recovery of the expected curve as the distribution of the sampled temperatures depends on it. One therefore wants to choose this prior so that the resulting distribution is flat or samples more the lower temperatures, for which the NQEs are more important. Here, we have tried different priors to obtain a satisfying distribution but this can be quite computationally expensive as multiple runs with different priors are needed. Note that to test different priors, a small number of live points $K$ can be used as we expect that the sampling frequency of the auxiliary distribution should be similar than with a higher value of $K$.

Overall, this trial-and-error methodology has been used throughout this paper to choose the parameters of the extended method: prior, sampling window $f$ and its parameter $\alpha$. This increases the computational cost of the analysis, as multiple choices are studied. This is especially true for the choice of the prior as a new exploration has to be performed for each prior considered. Changing the window or $\alpha$ only requires to repeat the post-processing step, which is much less computationally costly. In future works, to avoid this extra cost, a more systematic procedure to fine tune these parameters should be developed via an in-depth analysis of their influence on the extended partition function method in other systems.
\section{Conclusion}\label{sec_concl}
In this work, we have applied the nested sampling algorithm to evaluate the partition function with temperature-dependent effective potentials. Due to the explicit dependence on the temperature, the straightforward use of nested sampling implies to run distinct simulations for each temperature, which we refer to as the direct method. We have then proposed a method that allows to perform only one exploration for all temperatures, as in the classical case by focusing on the extended partition function. For this method, the partition function is computed over the configurational space as well as over an auxiliary temperature, corresponding to the temperature appearing in the potential. Hence, there is an extra parameter to sample compared to the direct method. However, even if one exploration of the extended partition function is more expensive than the exploration of the original partition function at a single temperature, the former requires less energy evaluations than the latter to compute the different thermodynamic properties. 

Specifically to this work, we considered temperature-dependent potentials resulting from the use of the path-integral formalism. The two methods have been compared on two systems: the harmonic potential and Lennard-Jones clusters. The direct method was first tested on the harmonic potential, for which \texttt{nested\_fit} was able to recover the expected analytical results. Moreover, it was used to study two types of Lennard-Jones clusters: Kr$_7$, which is almost classical, and Ne$_3$ and Ne$_{13}$, for which the nuclear quantum effects are important. For the latters, we have studied the convergence of the heat capacity with the number of live points $K$ and the number of replicas $P$. The number of live points $K$
increases swiftly with the number of replicas $P$.

We have seen that the extended method requires more live points than the direct method to reach convergence but has a smaller computational time, mostly due to only one exploration being done instead of several. Furthermore, we have seen that the extended method has additional parameters, such as the size of the regularized Dirac function $\alpha$ and the choice of the parameter (with uniform prior) to sample, that need to be finely tuned to each case studied. Among these additional parameters, some need only be chosen at the post-processing step ($\alpha$) while others need to be chosen before the nested sampling exploration. In the former case, the same exploration can be used for various values of the parameter. For the latter, a quick exploration can be performed with a reduced number of sampling points to select the optimal values.

In the future, the extended method with nested sampling could be applied to realistic systems as it has the advantage of requiring a single exploration to compute the thermodynamic properties. Furthermore, we envisage strategies for the fine tuning of the parameters of the extended method, which have a non-negligible impact on the recovery of the curves such as the prior distribution of the auxiliary temperature. Indeed, for the moment, we have used a trial-and-error methodology to choose the values of the parameters. Ideally, an automatic tuning of the parameters could be done. To achieve this, a more in-depth analysis of the method (both using exact results on test systems and recursive strategies) is likely needed in the future.


%
%

%


\section*{Supplementary Material}
The supplementary material includes the discrete expressions of the internal energy and heat capacity for a temperature-dependent potential for both the direct and extended methods; the sampling distribution of the temperature $\tilde{T}$ for the explorations used in Figure \mbox{\ref{fig_harm_q_method_ext}}; the heat capacity curves obtained for different value of $K$ for the Ne$_3$ case for both the direct and extended methods; the evolution of the MAPE with $K$ for different values of $\alpha$ for $P=4$ and $P=8$ for the Ne$_3$ case; the classical heat capacity for Ne$_{13}$, and a study of the Ne$_{13}$ case using the extended method with a uniform prior on $\tilde{\beta}$.

\begin{acknowledgments}
This work has been realized with the support of the Sorbonne Center for Artificial Intelligence—Sorbonne University—IDEX SUPER 11-IDEX-0004. L.M. thanks Livia Bartók-Pártay, Florent Calvo, Paola Cinella and Tony Lelièvre for their comments. L.M. thanks César Godinho for his work on \texttt{nested\_fit}. This work was granted access to the HPC resources of TGCC under the allocation A0130906719 made by GENCI. 
\end{acknowledgments}

\section*{Data Availability Statement}
The data that support the findings of this study are available from the corresponding author upon reasonable request.

\section*{Author declarations}
\subsection*{Conflict of interest}
The authors have no conflicts to disclose.
\subsection*{Author contributions}
\textbf{Philippe Depondt}: Conceptualization (supporting); methodology (supporting);  resources (equal); writing – review and editing (equal). \textbf{Fabio Finocchi}: Conceptualization (supporting); methodology (supporting);  supervision (supporting); resources (equal); writing – review and editing (equal). \textbf{Simon Huppert}: Conceptualization (supporting); methodology (supporting);  resources (equal); writing – review and editing (equal). \textbf{Lune Maillard}: Conceptualization (lead); data curation (lead); formal analysis (lead); investigation (lead); methodology (lead); software (supporting); validation (lead); visualization (lead); writing - original draft preparation (lead). \textbf{Thomas Plé}: Conceptualization (supporting);  methodology (supporting). \textbf{Julien Salomon}: Conceptualization (supporting); methodology (supporting);  supervision (supporting); writing – review and editing (equal). \textbf{Martino Trassinelli}: Conceptualization (supporting); data curation (supporting); methodology (supporting);  software (lead); supervision (lead); writing – review and editing (equal).

\appendix
\section{Nested sampling algorithm}\label{app_search}
\subsection{Approaching the density of states}
As presented in Section \ref{sssec_ns_pres}, at each iteration of the nested sampling algorithm, we 
replace the point $\boldsymbol{x}_{\mathrm{old}}$ by the point $\boldsymbol{x}_{\mathrm{new}}$ verifying $V(\boldsymbol{x}_{\mathrm{new}})<V(\boldsymbol{x}_{\mathrm{old}})$. Therefore, at each iteration $\ell$, we want to uniformly sample the region of space given by the energy constraint $V(\boldsymbol{x})<V(\boldsymbol{x}_{\mathrm{old}})$. We denote $\rho_\ell$ the cumulative DOS of this region. Since the points are uniformly distributed in the available space, their cumulative densities of states are uniformly distributed over $[0,1]$ \cite{speagle_dynesty_2020, ashton_nested_2022}. Therefore, $\rho_\ell/\rho_{\ell-1}$ follows a $\mbox{Beta}(K,1)$ distribution --- which corresponds to the distribution of the outermost value of a set of $K$ samples following a $\mbox{Uniform}([0,1])$ distribution. We thus estimate $\rho_\ell/\rho_{\ell-1}$ by its geometric mean \cite{skilling_nested_2004, skilling_nested_2006, ashton_nested_2022}, i.e., $\rho_\ell/\rho_{\ell-1}\approx \mathrm{e}^{-1/K}$. Therefore
\begin{equation}
\rho_\ell\approx \mathrm{e}^{-\ell/K}.    
\end{equation}
We therefore have that $w_\ell$, which approximates the DOS, is written
\begin{equation}\label{eq_weight_app}
w_\ell=\frac{1}{2}\left(\rho_{\ell-1}-\rho_{\ell+1}\right)=\frac{1}{2}\left(\mathrm{e}^{-\frac{\ell-1}{K}}-\mathrm{e}^{-\frac{\ell+1}{K}}\right).
 \end{equation}
\subsection{Searching new points}
\begin{figure}[h!]
    \centering
    \includegraphics[scale=1]{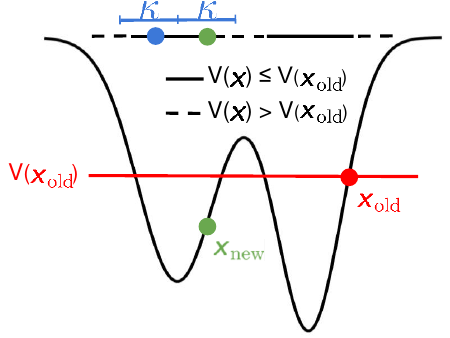}
    \caption{Representation of slice sampling in 1D. The potential $V$ is represented in gray and the constraint given by $V(\boldsymbol{x}_{old})$ by the red line. The point $\boldsymbol{x}_{\textrm{new}}$ is found by taking one of the live points (blue point) and building an interval around it (blue interval). A new point verifying the constraint is then sampled (green point).}
    \label{fig_slice_sampling}
\end{figure}\texttt{Nested\_fit}, which is the program used here, uses the slice sampling algorithm \cite{neal_slice_2003, handley_polychord_2015-1} to find the new point $\boldsymbol{x}_{\textrm{new}}$. This method consists in uniformly choosing new exploration points on a slice of the volume defined by the constraint $V(\boldsymbol{x})<V(\boldsymbol{x}_{\textrm{old}})$. In the one-dimensional case, represented in Figure \ref{fig_slice_sampling}, one of the live points (blue point in Figure \ref{fig_slice_sampling}) is randomly chosen and an interval, called slice, is built by randomly placing an interval of size $\kappa$ around it and then extending this interval on both sides --- by intervals of size $\kappa$ --- until its limits have an energy higher than $\boldsymbol{x}_{\textrm{old}}$ or are out of the sampling space (blue line in Figure \ref{fig_slice_sampling}). A point is then sampled from within the slice and accepted if it satisfies the constraint (green point in Figure \ref{fig_slice_sampling}) and rejected otherwise. In that case, another point is sampled until an acceptable point is found. In a multidimensional setting, a change of coordinates is first performed to efficiently explore all the parameter space, even in presence of strong correlation. This transformation is done via the Cholesky transformation of the covariance matrix of the live points of the considered step to transform the points coordinates into new coordinates with dimensions $\sim\mathcal{O}(1)$ in all directions \cite{handley_polychord_2015-1}. The one-dimensional algorithm is then applied recursively to the vectors of $n_{\textrm{bases}}$ randomly generated orthonormal bases\cite{maillard_assessing_2023}. Hence, $n_{\textrm{bases}}\times 3N$ steps are performed. The steps can be performed in the transformed space (\emph{Slice Sampling Transformed}) or in the real space (\emph{Slice Sampling Real})\cite{maillard_nested_2025}. In this work, we take $\kappa=1$ as it satisfies our requirements. 

\section{Discrete expressions for the path-integral internal energy and heat capacity}\label{app_PI_discrete_expressions}
\subsection{Direct method}
We denote $E_i(\beta_j)=V_P(\boldsymbol{x}^{(i)},\beta_j)$, $V_i=\bar{V}(\boldsymbol{x}^{(i)})$ and $Q_i(\beta_j)=Q(\boldsymbol{x}^{(i)},\beta_j)$, i.e. $E_i(\beta_j)=V_i+Q_i(\beta_j)$, $\boldsymbol{x}^{(i)}$ the position vector at iteration $i$ and $\beta_j=1/(k_\textrm{B}T_j)$ the inverse temperature we are considering. From Eqs. \eqref{eq_u_temp_dep}, \eqref{eq_cv_temp_dep}, \eqref{eq_nf_direct_average}, \eqref{eq_V_der_1} and \eqref{eq_V_der_2}, we obtain the following expressions for the internal energy and the heat capacity:

\begin{equation}\label{eq_u_nf_q}
U_\textrm{c}^P(\beta_j)=\frac{\sum_iw_i(V_i-Q_i(\beta_j))\mathrm{e}^{-\beta_j E_i(\beta_j)}}{Z_\textrm{c}^P(\beta_j)},   
\end{equation}
and
\begin{align}
C_\textrm{v,c}^P(\beta_j)=&\frac{\sum_iw_i(V_i-Q_i(\beta_j))^2\mathrm{e}^{-\beta_j E_i(\beta_j)}}{k_\textrm{B}T_j^2Z_\textrm{c}^P(\beta_j)}\nonumber\\
&-\frac{\left(\sum_iw_i(V_i-Q_i(\beta_j))\mathrm{e}^{-\beta_j E_i(\beta_j)}\right)^2}{k_\textrm{B}T_j^2Z_\textrm{c}^P(\beta_j)^2}\nonumber\\
&-2\frac{\sum_iw_iQ_i(\beta_j)\mathrm{e}^{-\beta_j E_i(\beta_j)}}{T_jZ_\textrm{c}^P(\beta_j)}\label{eq_cv_nf_q}
\end{align}
for an exploration performed at given inverse temperature $\beta_j$.

\subsection{Extended partition function method}
Within the extended partition function method, from Eqs. \eqref{eq_u_ext_method_exact}, \eqref{eq_cv_ext_method_exact}, \eqref{eq_nf_extended_average}, \eqref{eq_V_der_1} and \eqref{eq_V_der_2}, the contribution of $V_P$ to the internal energy and the heat capacity, respectively, read:
\begin{equation}
U_\textrm{c}^P(\beta)\approx\frac{\sum_i w_i (V_i-Q_i) f(\beta-\tilde{\beta}_i;\alpha)\mathrm{e}^{-\beta E_i}}{Z_\textrm{c}^P(\beta)}
\end{equation}
and
\begin{eqnarray}
C_\textrm{v,c}^P(\beta)&\approx&-\frac{\left(\sum_i w_i (V_i-Q_i) f(\beta-\tilde{\beta}_i;\alpha)\mathrm{e}^{-\beta E_i}\right)^2}{k_\textrm{B} T^2 Z_\textrm{c}^P(\beta)^2}\nonumber\\
&&+\frac{\sum_i w_i (V_i-Q_i)^2 f(\beta-\tilde{\beta}_i;\alpha)\mathrm{e}^{-\beta E_i}}{k_\textrm{B} T^2 Z_\textrm{c}^P(\beta)}\nonumber\\
&&-2\frac{\sum_i w_i Q_i f(\beta-\tilde{\beta}_i;\alpha)\mathrm{e}^{-\beta E_i}}{T Z_\textrm{c}^P(\beta)},
\end{eqnarray}
with $w_i$ as in Eq. \eqref{eq_weight} and $\tilde{\beta}_i$ and $E_i=V_i+Q_i$ the inverse temperature and energy sampled at iteration $i$. The terms $E_i=E(\boldsymbol{x}^{(i)},\tilde{\beta}_i)$ and $Q_i=Q(\boldsymbol{x}^{(i)},\tilde{\beta}_i)$ are computed at inverse temperature $\tilde{\beta}_i$ for the position vector $\boldsymbol{x}^{(i)}$.

\section{Derivation of the exact expressions for the path-integral harmonic potential}\label{app_harm_P_der}
In this appendix, we derive the exact expressions for the heat capacity of the path-integral harmonic potential with $P=2$ and $P=4$. Since all dimensions are separable in Eq. \eqref{eq_VP_harm}, we compute the expressions for one particle in one dimension and multiply the results by $3$. We do not use reduced units in this section.
\subsection{Exact case with $P=2$ replicas}
Making the change of variables $y_1=\boldsymbol{x}_1-\boldsymbol{x}_2$ and $y_2=\boldsymbol{x}_1+\boldsymbol{x}_2$, we have that
\begin{align}
V_2(\boldsymbol{x}_1,\boldsymbol{x}_2,\beta)&=\frac{2m}{\hbar^2\beta^2}(\boldsymbol{x}_1-\boldsymbol{x}_2)^2+\frac{m\omega^2}{4}(\boldsymbol{x}_1^2+\boldsymbol{x}_2^2)\\
&=\left(\frac{2m}{\hbar^2\beta^2}+\frac{m\omega^2}{8}\right)y_1^2+\frac{m\omega^2}{8}y_2^2\\
&=V_2(y_1,y_2,\beta).   
\end{align}

Therefore, the contribution of $V_2$ to the partition function (Eq. \eqref{eq_zc_q_vp}) is:
\begin{align}
Z&_\textrm{c}^{P=2}(\beta)\nonumber\\
&=\iint_{\mathbb{R}^2} d\boldsymbol{x}_1d\boldsymbol{x}_2 \mathrm{e}^{-\beta V_2(\boldsymbol{x}_1,\boldsymbol{x}_2,\beta)}\\
&=\frac{1}{2}\iint_{\mathbb{R}^2} dy_1dy_2 \exp \left(-\beta \left(\left(\frac{2m}{\hbar^2\beta^2}+\frac{m\omega^2}{8}\right)y_1^2+\frac{m\omega^2}{8}y_2^2\right)\right)\\
&=\frac{1}{2}\sqrt{\frac{\pi}{\beta\left(\frac{2m}{\hbar^2\beta^2}+\frac{m\omega^2}{8}\right)}}\sqrt{\frac{8\pi}{\beta m\omega^2}}.
\end{align}
Consequently, we have
\begin{equation}
C_{\textrm{v,c}}^{P=2}(\beta)=k_\textrm{B}-\frac{4\frac{(k_\textrm{B}T)^4}{\hbar^4}+\frac{3}{4}\frac{(k_\textrm{B}T)^2\omega^2}{\hbar^2}}{\left(\frac{(k_\textrm{B}T)^2}{\hbar^2}+\frac{\omega^2}{8}\right)^2}k_\textrm{B}    
\end{equation}
and
\begin{equation}
C_\textrm{v}^{P=2}(\beta)=2k_\textrm{B}-\frac{4\frac{(k_\textrm{B}T)^4}{(\hbar\omega)^4}+\frac{3}{4}\frac{(k_\textrm{B}T)^2}{(\hbar\omega)^2}}{\left(2\frac{(k_\textrm{B}T)^2}{(\hbar\omega)^2}+\frac{1}{8}\right)^2}k_\textrm{B}.
\end{equation}
\subsection{Exact case with $P=4$ replicas}
With the change of variables $y_1=\boldsymbol{x}_1-\boldsymbol{x}_2-\boldsymbol{x}_3+\boldsymbol{x}_4$, $y_2=\boldsymbol{x}_1+\boldsymbol{x}_2-\boldsymbol{x}_3-\boldsymbol{x}_4$, $y_3=\boldsymbol{x}_1-\boldsymbol{x}_2+\boldsymbol{x}_3-\boldsymbol{x}_4$ and $y_4=\boldsymbol{x}_1+\boldsymbol{x}_2+\boldsymbol{x}_3+\boldsymbol{x}_4$, we have that
\begin{align}
V_4&(\boldsymbol{x}_1,\boldsymbol{x}_2,\boldsymbol{x}_3,\boldsymbol{x}_4,\beta)\nonumber\\
=&\frac{2m}{\hbar^2\beta^2}((\boldsymbol{x}_1-\boldsymbol{x}_2)^2+(\boldsymbol{x}_2-\boldsymbol{x}_3)^2+(\boldsymbol{x}_3-\boldsymbol{x}_4)^2+(\boldsymbol{x}_4-\boldsymbol{x}_1)^2)\nonumber\\
&+\frac{m\omega^2}{8}(\boldsymbol{x}_1^2+\boldsymbol{x}_2^2+\boldsymbol{x}_3^2+\boldsymbol{x}_4^2)\\
=&\left(\frac{m}{\hbar^2\beta^2}+\frac{m\omega^2}{32}\right)y_1^2+\left(\frac{m}{\hbar^2\beta^2}+\frac{m\omega^2}{32}\right)y_2^2\nonumber\\
&+\left(\frac{2m}{\hbar^2\beta^2}+\frac{m\omega^2}{32}\right)y_3^2+\frac{m\omega^2}{32}y_4^2\\
=&V_4(y_1,y_2,y_3,y_4,\beta).  
\end{align}

Hence
\begin{equation}
Z_\textrm{c}^{P=4}(\beta)=C'\sqrt{\frac{1}{\beta^4\left(\frac{1}{\hbar^2\beta^2}+\frac{\omega^2}{32}\right)\left(\frac{1}{\hbar^2\beta^2}+\frac{\omega^2}{32}\right)\left(\frac{2}{\hbar^2\beta^2}+\frac{\omega^2}{32}\right)}},
\end{equation}
where $C'$ is a constant that is independent of $\beta$.
Furthermore, 
\begin{align}
C_{\textrm{v,c}}^{P=4}(\beta)=2k_\textrm{B}&-2\frac{\frac{(k_\textrm{B}T)^4}{\hbar^4}+\frac{3}{32}\frac{(k_\textrm{B}T)^2\omega^2}{\hbar^2}}{\left(\frac{(k_\textrm{B}T)^2}{\hbar^2}+\frac{\omega^2}{32}\right)^2}k_\textrm{B}\nonumber\\
&-\frac{4\frac{(k_\textrm{B}T)^4}{\hbar^4}+\frac{3}{16}\frac{(k_\textrm{B}T)^2}{\hbar^2\omega^2}}{\left(2\frac{(k_\textrm{B}T)^2}{\hbar^2}+\frac{\omega^2}{32}\right)^2}k_\textrm{B}
\end{align}
and
\begin{align}
C_v^{P=4}(\beta)=4k_\textrm{B}&-2\frac{\frac{(k_\textrm{B}T)^4}{(\hbar\omega)^4}+\frac{3}{32}\frac{(k_\textrm{B}T)^2}{(\hbar\omega)^2}}{\left(\frac{(k_\textrm{B}T)^2}{(\hbar\omega)^2}+\frac{1}{32}\right)^2}k_\textrm{B}\nonumber\\
&-\frac{4\frac{(k_\textrm{B}T)^4}{(\hbar\omega)^4}+\frac{3}{16}\frac{(k_\textrm{B}T)^2}{(\hbar\omega)^2}}{\left(2\frac{(k_\textrm{B}T)^2}{(\hbar\omega)^2}+\frac{1}{32}\right)^2}k_\textrm{B}.
\end{align}

\section{Computational details}\label{app_comp_details}
The parameters used for the different applications presented in this work are given in Tables \ref{tab_par_harm_direct} for the harmonic potential with the direct method, \ref{tab_par_harm_ext} for the harmonic potential with the extended method, \ref{tab_par_lj_direct} for Lennard-Jones clusters with the direct method and \ref{tab_par_LJ_ext} for Lennard-Jones clusters with the extended method.
\begin{table}[h!]
\caption{\label{tab_par_harm_direct} Computational parameters for the quantum harmonic potential with the direct method.}
\begin{ruledtabular}
\begin{tabular}{lccc}
& $K$ & $n_{\textrm{bases}}$ & $1/(k_\textrm{B}\beta_\textrm{s})$ ($k_\textrm{B}T/(\hbar\omega)$)\\
\hline
Evolution & \multirow{2}{*}{1000} & \multirow{2}{*}{1} & \multirow{2}{*}{0.01 } \\
with $P$ & & &\\
Comparison & \multirow{2}{*}{25000} & \multirow{2}{*}{2} & \multirow{2}{*}{0.01} \\
with extended method & & &\\
\end{tabular}
\end{ruledtabular}
\end{table}
\begin{table}[h!]
\caption{\label{tab_par_harm_ext} Computational parameters for the quantum harmonic potential with the extended method.}
\begin{ruledtabular}
\begin{tabular}{lccccc}
& \multirow{2}{*}{$K$} & \multirow{2}{*}{$n_{\textrm{bases}}$} & $1/(k_\textrm{B}\beta_\textrm{s})$ &  Smeared-delta & Explored \\
& & & ($k_\textrm{B}T/(\hbar\omega))$ & function + $\alpha$ & parameter\\
\hline
Choosing & See & \multirow{2}{*}{2} & \multirow{2}{*}{0.01} &  See & \multirow{2}{*}{$1/\tilde{\beta}$}\\
$\alpha$ and $f$ & text & & & text & \\
Evolution & See & \multirow{2}{*}{2} & \multirow{2}{*}{0.01} &  Gaussian & See \\
with $P$ & text & &  & $\alpha\to$ see text & text \\
Comparison  & \multirow{3}{*}{50000} & \multirow{3}{*}{2} & \multirow{3}{*}{0.01} & \multirow{2}{*}{Gaussian} & \multirow{3}{*}{$1/\tilde{\beta}$} \\
with direct & & & & \multirow{2}{*}{$\alpha=7.1$} &\\
method & & & & & \\

\end{tabular}
\end{ruledtabular}
\end{table}
\begin{table}[h!]
\caption{\label{tab_par_lj_direct} Computational parameters for the quantum Lennard-Jones clusters with the direct method.}
\begin{ruledtabular}
\begin{tabular}{lccc}
 & $K$ & $n_{\textrm{bases}}$ & $1/(k_\textrm{B}\beta_\textrm{s})$ ($k_\textrm{B}T/\epsilon$)\\
\hline
Ne$_3$& See text & 5 & 0.01 \\
Kr$_7$& 2000 ($P=1$), 5000 ($P=2$) & 4 & 0.01\\
Ne$_{13}$ & 16000 & 3 & 0.01\\
\end{tabular}
\end{ruledtabular}
\end{table}
\begin{table}[h!]
\caption{\label{tab_par_LJ_ext} Computational parameters for the quantum Lennard-Jones clusters with the extended method.}
\begin{ruledtabular}
\begin{tabular}{lccccc}
& \multirow{2}{*}{$K$} & \multirow{2}{*}{$n_{\textrm{bases}}$} & $1/(k_\textrm{B}\beta_\textrm{s})$ &  Smeared-delta & Explored\\
& & & ($k_\textrm{B}T/\epsilon$) & function + $\alpha$ & parameter\\
\hline
\multirow{3}{*}{Ne$_3$} &  \multirow{3}{*}{See text} & \multirow{3}{*}{2} & \multirow{3}{*}{0.01 } & Gaussian & \multirow{3}{*}{$\tilde{\beta}$} \\
&  & & & $\alpha=1$ ($P=2,4$) & \\
&  & & & $\alpha=1.4$ ($P=8,16$) & \\
Ne$_{3}$ & \multirow{2}{*}{See text} & \multirow{2}{*}{5} & \multirow{2}{*}{0.01} & \multirow{2}{*}{/} & \multirow{2}{*}{$\tilde{\beta}$} \\
(Table \ref{tab_tps_LJ_Q_methods}) & & & & & \\
\multirow{2}{*}{Ne$_{13}$} & \multirow{2}{*}{See text} & \multirow{2}{*}{3} & \multirow{2}{*}{0.01} & Gaussian & \multirow{2}{*}{$\tilde{\beta}^8$} \\
& & & & $\alpha=1$ & \\
\end{tabular}
\end{ruledtabular}
\end{table}
\section{Choice of the box size}\label{app_box_size}
When presenting the change of variables in Section \ref{ssec_extended_method}, we mentioned that it makes the choice of the boundaries given to the replicas ($\Tilde{\boldsymbol{y}}_i$, $1\leq i \leq P-1$) in \texttt{nested\_fit} more complex. There are multiple ways to choose those boundaries, noting $a$ the size of the interval:
\begin{enumerate}
    \item Choosing $a$ so that it corresponds to the size of the box in real space, i.e., $a\lambda_P=L$.
    \item In a similar way, choosing $a$ so that $a\lambda_{\textrm{th}}=L$ where 
    \begin{equation}\label{eq_lambda_th}
        \lambda_{\textrm{th}}=\sqrt{\frac{2\pi\hbar^2\beta}{m}}
    \end{equation}
    is the de Broglie thermal wavelength which gives information about the delocalization of the particle \cite{silvera_boseeinstein_1997}.
    \item If we look at the transformed potential, we have that the contribution to the partition function of the interaction between replicas is $\exp\left(-\frac{\beta}{2}(\Tilde{\boldsymbol{y}}_i-\Tilde{\boldsymbol{y}}_{i+1})^2\right)$ which is a Gaussian function with standard deviation
    \begin{equation}\label{equ_sigma_limit}
        \sigma=\sqrt{\frac{1}{\beta}}.
    \end{equation}
    Hence, we choose $L$ to be between $3\sigma$ to $5\sigma$.
\end{enumerate}

\begin{figure}[h!]
    \centering
    \includegraphics[scale=1,trim=0 12 0 10,clip]{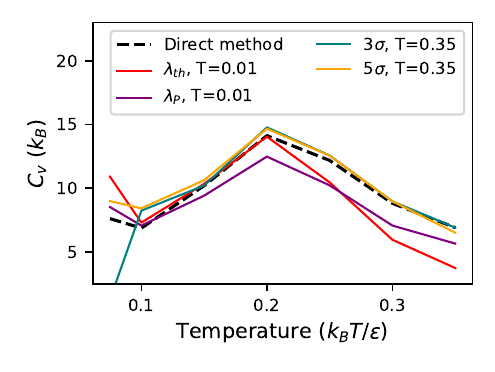}
    \caption{Heat capacity for the 3-atoms Neon cluster with $P=8$ replicas and $K=32768$ live points. Comparison of the different choices for the size of the box. The reference curve in dashed blue was obtained with the direct method with the highest $K$. We do not consider temperature below $0.075\frac{k_\textrm{B}T}{\epsilon}$ as they are not converged with $P$. The quantities $\lambda_{\textrm{th}}$, $\lambda_P$ and $\sigma$ are defined in Eqs. \eqref{eq_lambda_th}, \eqref{eq_lambda_p} and \eqref{equ_sigma_limit}, respectively.}
    \label{fig_lj_q_limits_p_8}
\end{figure}

In the three cases, the parameters depend on the temperature. In the first two cases, the box is bigger for low temperatures and smaller for high temperatures at constant $a$: we choose $a$ at the lowest temperature, so that the replicas always stay in the box at any temperature. For the last case, on the contrary, the box is bigger for high temperatures than for low temperatures; we thus take the box corresponding to the highest temperatures. Figure \ref{fig_lj_q_limits_p_8} presents the curves obtained for the $P=8$ case using $\lambda_{\textrm{th}}$ (Eq. \eqref{eq_lambda_th}) and $\lambda_P$ (Eq. \eqref{eq_lambda_p}) with $T=0.01k_\textrm{B}T/\epsilon$, and 3 and 5 times the Gaussian standard deviation with $T=0.35k_\textrm{B}T/\epsilon$. We clearly see that, when we take the lowest temperature with $\lambda_P$ and $\lambda_{\textrm{th}}$, the heat capacity is not correctly recovered for the higher range of temperatures, the volume explored in that case probably not correctly representing the energy surface. This effect is less visible for lower $P$ as the parameter space is smaller. Choosing the Gaussian with $3\sigma$ or $5\sigma$ recovers the reference curve quite well. The lowest temperature point is of little interest as it is at a non-converged temperature (in terms of $P$). Hence, we use a box of size $3\sigma$ or $5\sigma$. For $P=2$ and $P=4$, we use $\lambda_{\textrm{th}}$ with $T=0.01k_\textrm{B}T/\epsilon$, as this choice only becomes visibly inadequate for $P\geq8$, for which we use a $3\sigma$ box.

\bibliography{Biblio_QNest}
\end{document}